\documentclass[onecolumn,authoryear]{els-mrw} 

\usepackage{amsmath,amssymb,amsfonts,amsthm,makeidx,graphicx}
\usepackage{txfonts}
\usepackage{helvet}
\usepackage[colorlinks=true,citecolor=blue,filecolor=blue,linkcolor=black,urlcolor=blue,pdftex]{hyperref}
\usepackage{caption}
\usepackage{subcaption}

\def\aj{AJ}%
\def\araa{ARA\&A}%
\def\apj{ApJ}%
\def\apjl{ApJ}%
\def\apjs{ApJS}%
\def\aap{A\&A}%
\def\aapr{A\&A~Rev.}%
\def\aaps{A\&AS}%
\def\mnras{MNRAS}%
\def\prd{Phys.~Rev.~D}%
\def\pasp{PASP}%
\def\ssr{Space~Sci.~Rev.}%
\def\nat{Nature}%
\def\iaucirc{IAU~Circ.}%
\def\nphysa{Nucl.~Phys.~A}%
\def\physrep{Phys.~Rep.}%
\newcommand{\pasa}{Publ.\ Astron.\ Soc.\ Aust.\ }

\newcommand{\blue}[1]{\textcolor{black}{{#1}}}
\newcommand{\bluee}[1]{\textcolor{black}{{#1}}}

\begin{document}

\chapter{Core-collapse supernovae}\label{chap1}

\author[1]{Anders Jerkstrand}%
\author[2]{Dan Milisavjlevic}%
\author[3]{Bernhard Müller (alphabetical)}%

\address[1]{\orgname{Stockholm University}, \orgdiv{Department of Astronomy}, \orgaddress{Roslagstullsbacken 21, 11421 Stockholm, Sweden.}}
\address[2]{\orgname{Purdue University}, \orgdiv{Department of Physics and Astronomy}, \orgaddress{525 Northwestern Ave., West Lafayette, IN 47907, USA.}}
\address[3]{\orgname{Monash University}, \orgdiv{School of Physics and Astronomy}, \orgaddress{10 College Walk, Clayton, VIC 3800, Australia.}}

\articletag{Chapter Article tagline: update of previous edition,, reprint..}

\maketitle


\begin{glossary}[Nomenclature]
\begin{tabular}{@{}lp{34pc}@{}}
SN & Supernova\\
CCSN & Core-collapse supernova\\
SNR & Supernova remnant\\
ZAMS & Zero-Age Main Sequence\\
RSG & Red Supergiant\\
BSG & Blue Supergiant \\
WR & Wolf-Rayet star\\
NS & Neutron star\\
BH & Black hole\\
GRB & Gamma-Ray Burst\\
PWN & Pulsar Wind Nebula\\
CSM & Circumstellar Medium\\
CSI & Circumstellar interaction\\
SESN & Stripped-envelope supernova \\
LTE & Local Thermodynamic Equilibrium\\
NLTE & Non-Local Thermodynamic Equilibrium\\
\end{tabular}
\end{glossary}

\begin{abstract}[Abstract]
Core-collapse supernovae (CCSNe) are the explosive end-points of stellar evolution for $M_{ZAMS} \gtrsim 8$ $M_\odot$ stars. The cores of these stars collapse to neutron stars, a process in which high neutrino luminosity drives off the overlying stellar layers, which get ejected with thousands of kilometers per second. These supernovae enrich their host galaxies with elements made both during the star's life and in the explosion, providing the main cosmic source of elements such as oxygen, neon and silicon. Their high luminosities ($\sim$ $10^{42}$ erg s$^{-1}$ at peak) make SNe beacons to large distances, and their light curves and spectra provide rich information on single and binary stellar evolution, nucleosynthesis, and a diverse set of high-energy physical processes. As the SN ejecta sweep up circumstellar and interstellar matter, it eventually enters a supernova remnant phase, exemplified by nearby, spatially resolved remnants such as Cas A and the Crab Nebula. In this phase, shocks and pulsar winds continue to light up the interior of the exploded stars, giving detailed information about their 3D structure. We review the central concepts of CCSNe, from the late stages of evolution of massive stars, through collapse, explosion, and electromagnetic display, to the final remnant phase. We briefly discuss still open questions, and current and future research avenues.
\end{abstract}

\begin{BoxTypeA}[chap1:box1]{Key points}
\begin{itemize}
    \item \textbf{Core-collapse supernovae} occur when the cores of massive stars collapse to neutron stars, blowing off the outer layers.
    \item \textbf{Neutrinos} play a key role in transferring the gravitationally released energy of the core to kinetic energy of the mantle.
    \item \textbf{Supernova light curves and spectra} inform us about the mass and composition of the exploded star. CCSNe are the main cosmic source of most of the elements in the atomic number range $Z=8-37$.
    \item \textbf{Supernova remnants} allow spatially resolved views of the debris and the central compact objects (typically pulsars) left behind.
\end{itemize}
\end{BoxTypeA}

\section{Introduction}\label{chap1:sec1}

Core-collapse supernovae (CCSNe) are the explosions of massive stars, $M_{ZAMS} \gtrsim 8\ M_\odot$, as they reach the end of their evolution. Such stars form iron cores at the end of their lives (apart from a small range of stars, $M_{ZAMS} \sim 8-9$ $M_\odot$, that may explode already at the ONeMg core stage), which eventually become unstable and collapse to neutron stars (NSs) or black holes (BHs), a process in which the overlying stellar layers get violently expelled. In this way, the cosmos is enriched in elements, with galactic chemical evolution models predicting CCSNe to dominate production of elements with atomic numbers $Z=8-23,29-37$, and possibly several of the yet heavier elements \citep{Timmes1995,Kobayashi2020}. In addition,  compact objects (NSs and BHs) are formed, some of which lead to gamma-ray bursts \citep{Woosley2006} and NS mergers \citep{Abbott2017}, and energy and momentum are injected into the surroundings triggering large-scale mass motions and regulating star formation \citep{colling_18}. 

One can estimate that about one CCSN occurs per second somewhere in the Universe, with a relatively good match between star formation rates for massive stars and CCSN rates \citep{Boticella2012}. CCSN progenitors live short lives and these SNe therefore occur in star-forming galaxies \citep{Schulze2021}. We currently (2025) discover about about one in every thousand CCSN, or 10,000 per year. The discovery rate has dramatically increased over the past two decades thanks to automated wide-field transient surveys such as Pan-STARRS, The Lick Observatory Supernova Survey,  The Supernova SkyMapper, and the Zwicky Transient Facility (ZTF), and will soon take another major leap with the development of the Vera C.\ Rubin Observatory. Thanks to their immense brightness ($\sim 10^9\ L_\odot$ at peak) we can observe SNe to large distances (hundreds of Mpc), but they are also common enough (about 2 per century in a Milky Way-like galaxy) that we can regularly study them in nearby galaxies (to several tens of Mpc, and sometimes yet further), and up-close in spatially resolved remnants in the Milky Way (typical distance a few kpc).\\

\begin{figure}[tbp]
\centering
\begin{subfigure}{0.49\textwidth}
\includegraphics[width=1\linewidth]{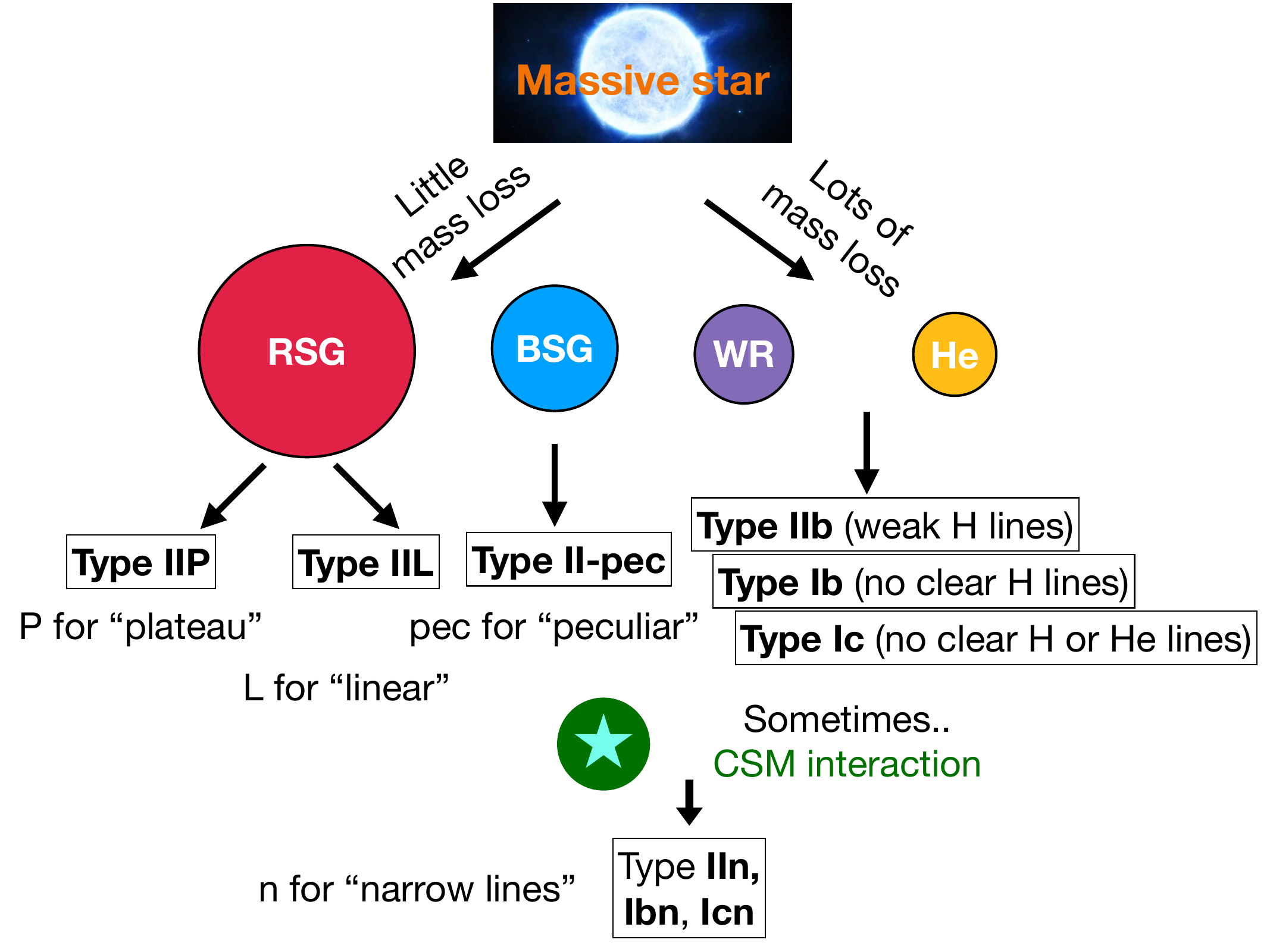}
\end{subfigure}
\begin{subfigure}{0.49\textwidth}
\includegraphics[width=0.95\linewidth]{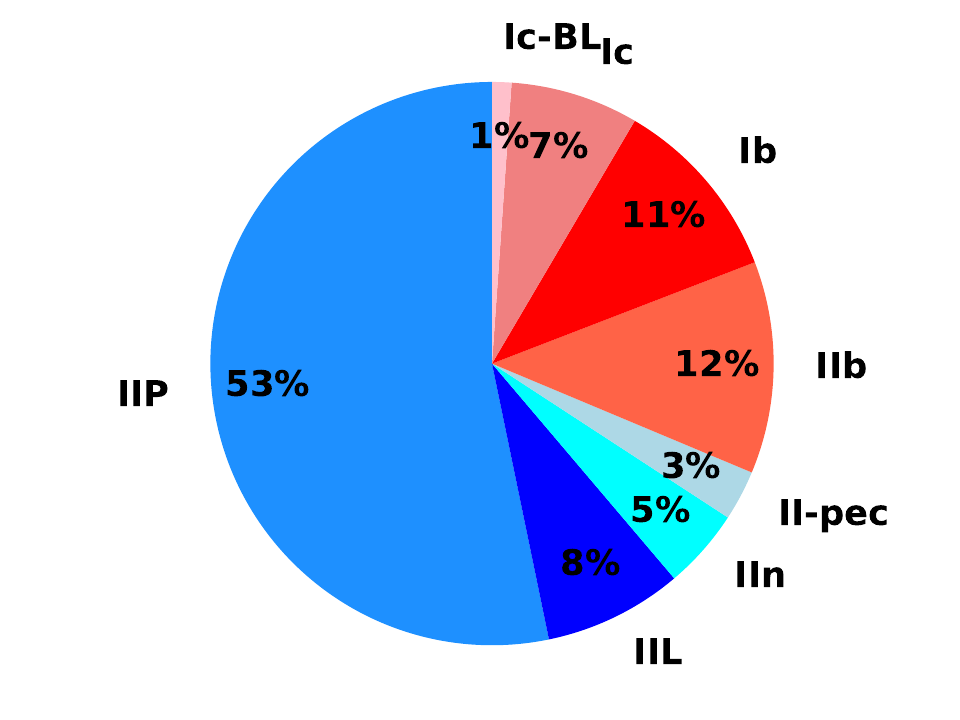}
\end{subfigure}
\caption{\emph{Left:} Progenitors and classification scheme of CCSNe. \emph{Right:} Observed fractions of CCSN types in a volume-limited sample.
Data from \citet{Li2011} and rates from \citet{Shivvers2017}. Stars in the blue-tinted classes (about 2/3 of the total) have lost relatively little material over their lives, by winds and binary mass transfer, whereas those in the the red-tinted (about 1/3 of the total) have lost significant.}
\label{fig:class-scheme}
\end{figure}

\begin{figure}[tbp]
\includegraphics[width=1\linewidth]{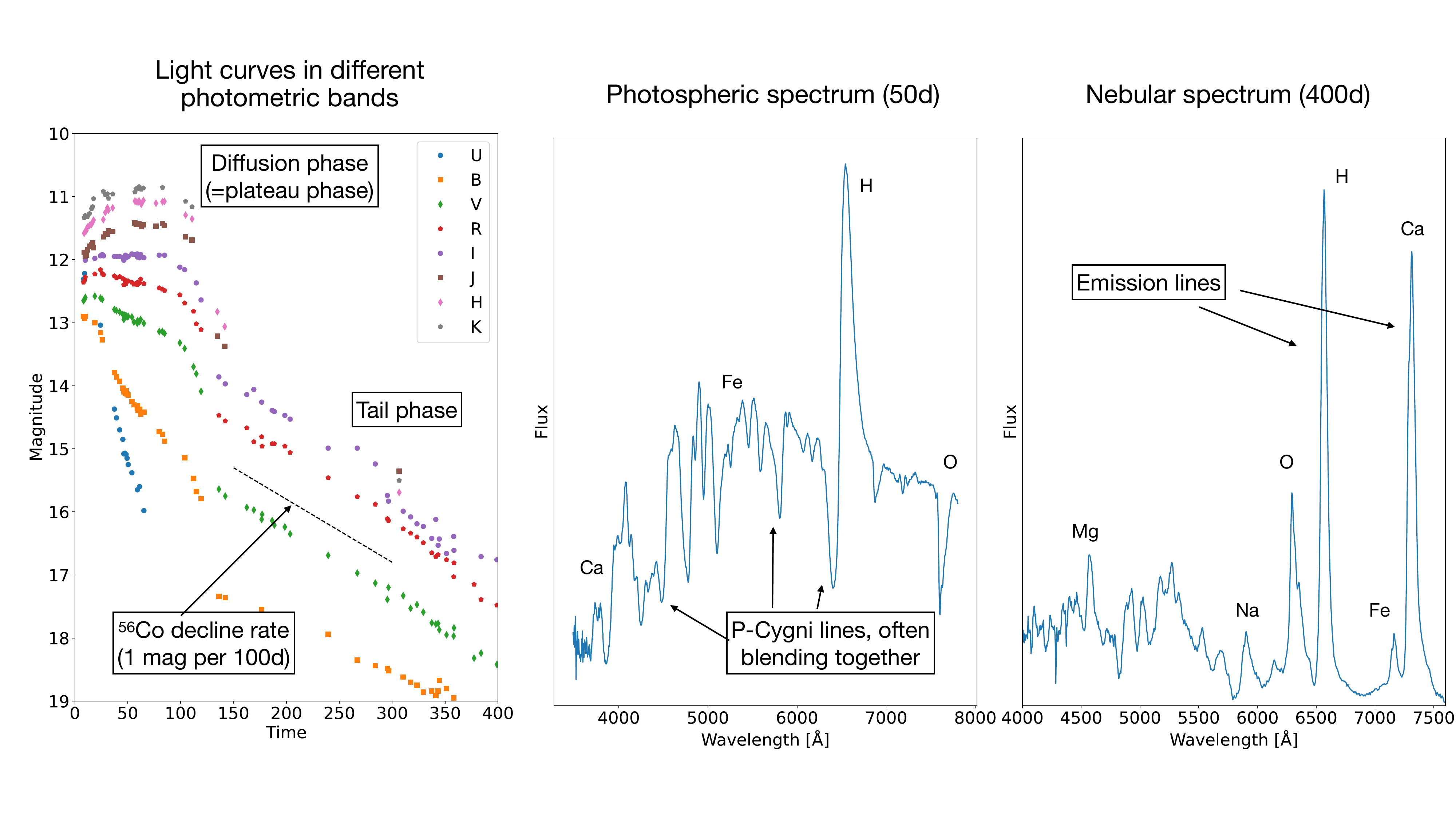}
\caption{Light curves in different photometric bands (left), photospheric spectrum (middle), and nebular spectrum (right) of a typical Type IIP SN. Data from \citet{Maguire2010}.}
\label{fig:examples}
\end{figure}

\noindent \textbf{Classification system.} The empirical classification of CCSNe \citep[see][for overviews]{Filippenko1997,GalYam2017,Modjaz2019} initially divides into the branches of Type I (hydrogen lines absent) and Type II (hydrogen lines present) - with \citet{Minkowski1941} generally being credited with taking this first step for the taxonomy. The Type I class subdivides into Type Ib (helium lines present), and Type Ic (helium lines absent). Some Type Ic SNe have very broad lines, defining a specific Type Ic-BL class. There is also a Type Ia class, which are not CCSNe but thermonuclear explosions of white dwarfs. The distinction between the different Type I SN classes emerged in the mid 1980s \citep[e.g.][]{Wheeler1985}. The Type II branch subdivides into Type IIP (the light curve shows a plateau), Type IIL (the light curve shows a linear decline), Type II-pec (the light curve has a long initial rise) and Type IIb (spectra show weak H lines only, that later disappear). If narrow lines are seen, produced by the surrounding, slow circumstellar gas rather than the SN ejecta (which emit broad lines due to the high expansion velocity), the classification becomes Type IIn \citep{Schlegel1990}, Ibn \citep{Pastorello2008}, and Icn \citep{GalYam2021,Perley2022} respectively, depending on the composition of the circumstellar material (CSM).  Figure \ref{fig:class-scheme} summarizes the classification system (left), and 
shows observationally inferred rates of the different classes (right). The most common type of supernova can be seen to be Type IIP, which make up just over half of all CCSNe by unit volume. An example of the light curves and spectra of a typical Type IIP SN (SN 2004et) is shown in Figure \ref{fig:examples}. Note the pronounced plateaus on the optical ($V,R,I$) bands, whereas UV/blue bands ($U,B$) decline and the infrared bands ($J,H,K$) increase over the diffusion phase ($\sim 0-100$d).

It is today understood that the Type IIP/L-IIb-Ib-Ic differentiation arises due to different degrees of \emph{mass loss} experienced by the star over its life up to explosion. Type IIb/Ib/Ic SNe are collectively called stripped-envelope supernovae (SESNe). This mass loss may occur both by stellar winds and by gravitational mass transfer to a companion star.  Both these processes are mainly important in the post-main sequence phase when massive stars expand significantly, boosting their wind mass loss rates due to reduced surface gravity, as well as the chances of making Roche lobe contact with a binary companion. For a long time were Wolf-Rayet stars seen as the plausible progenitors of Type Ib/c SNe - however evidence has accumulated that most of these SNe instead appear to arise from lower-mass binary-stripped stars. There has historically been few known observational counterparts of these progenitors stars, as their bright companion stars tend to mask their light, but recently new discoveries have been reported \citep{Drout2023}. One should also note that many Type II SN progenitors may have gained mass from a companion \citep{Zapartas2019}.  Mass loss again manifests itself for the Type IIn, Ibn and Icn classes, believed to arise when the SN ejecta collide with (unusually) dense circumstellar material.

\label{fig:Li2011}

\section{Progenitors and explosion}
\label{sec:bernhard}
\subsection{Progenitor evolution up to collapse}
\begin{figure}
    \centering
    \includegraphics[width=0.6\linewidth]{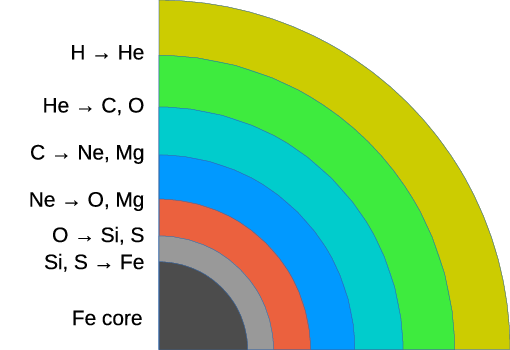}
    \caption{Sketch (not to scale) of the onion shell structure of a massive star prior to the collapse of the iron core, in this case a red supergiant (RSG) progenitor with a hydrogen envelope.  
    Inside the hydrogen envelope, there are shells of the ashes of hydrogen, helium, carbon, neon, and oxygen burning, and finally the iron core built by silicon burning. The respective burning processes are taking place at the bottom of the shells, but note that some burning shells may not be active at the time of collapse. The major fuels and ashes for each burning stage are indicated on the left.}
    \label{fig:onion}
\end{figure}

The progenitors of core-collapse supernovae are \emph{massive stars} with main sequence masses between about $8\ M_\odot$ and $140\  M_\odot$ if these evolved as single stars
. For stars in binary systems the initial masses may be different due to mass exchange with their companions
, and the lower and upper boundary for single stars or stars that have undergone mass transfer are subject to uncertainties in stellar physics. Such massive stars go through a sequence of six distinct burning stages (H, He, C, Ne, O, and Si burning) in the core and eventually develop an iron core after Si burning \citep{Woosley2002,woosley_05}. Burning of lighter elements proceeds in shells outside the core and leaves the star with an ``onion-shell'' structure (Figure~\ref{fig:onion}). The iron core has low entropy and is therefore mostly supported against gravity by (relativistic) electron degeneracy pressure, 
i.e., the pressure exerted by
electrons as they occupy states of non-zero momentum and kinetic energy (up to about the \emph{Fermi energy})
due to the Pauli exclusion principle.
In addition to the electron degeneracy pressure, thermal radiation pressure also contributes, especially in more massive core-collapse supernova progenitors. As Si shell burning adds more mass to the iron core, it contracts and eventually reaches densities $\rho \gtrsim 10^{9}\,\mathrm{g}\,\mathrm{cm}^{-3}$.

At such high densities, the high Fermi energy of electrons allows electron captures on heavy nuclei, which drains the degeneracy pressure and therefore accelerates contraction. In addition, temperatures in the core become high enough for some measure of photodisintegration of heavy nuclei, which reduces radiation pressure support. Contraction thus becomes a runaway process and accelerates to a collapse on a dynamical time scale $\tau_\mathrm{dyn}\sim (G\rho)^{-1/2}$, or a few hundred $\mathrm{ms}$. The onset of collapse occurs when the iron core has grown to about
$1.4\ M_\odot$, corresponding roughly to the limiting Chandrasekhar mass of slightly neutron-rich matter. At this point, the core has reached  central densities of a few $10^{9}\,\mathrm{g}\,\mathrm{cm}^{-3}$, central temperatures slightly below $10^{10}\,\mathrm{K}$, and has a radius of $\gtrsim 1000\,\mathrm{km}$ \citep{Woosley2002}. In a narrow regime between low-mass and high-mass stars, where degenerate conditions in the core are encountered already after C burning, collapse may be initiated earlier by electron captures on Ne and Mg, giving rise to \emph{electron-capture supernovae}
\citep{Nomoto87,leung_19}. Such electron-capture supernovae may explain some observed low-energy explosions, but details of this evolutionary channel are still debated.

\subsection{Collapse and bounce}
Electron captures proceed at an accelerating pace until the
the core reaches densities of $\mathord{\sim}10^{12}\,\mathrm{g}\,\mathrm{cm}^{-3}$. At this point, neutrinos from electron capture reactions become trapped and hence further loss of leptons from the core stops,
but this is not sufficient to halt the collapse.
The collapse eventually stops when the core exceeds nuclear saturation density ($\mathord{\sim}2.7\times 10^{14}\,\mathrm{g}\,\mathrm{cm}^{-3}$). At this point, matter becomes significantly more incompressible due to repulsive nuclear interactions, and the inner core rebounds elastically (``bounce''). This results in the launching of a shock wave into the outer part of the core that is still collapsing supersonically. As the shock dissociates the infalling iron-group material into free protons and neutrons and rapid neutrino losses occur once the post-shock densities drop below $\mathord{\sim}10^{12}\,\mathrm{g}\,\mathrm{cm}^{-3}$, the shock ``stalls'' and turns into an accretion shock within milliseconds \citep{Bethe1990,Janka2007}.

\begin{figure}
    \centering
    \includegraphics[width=0.6\linewidth]{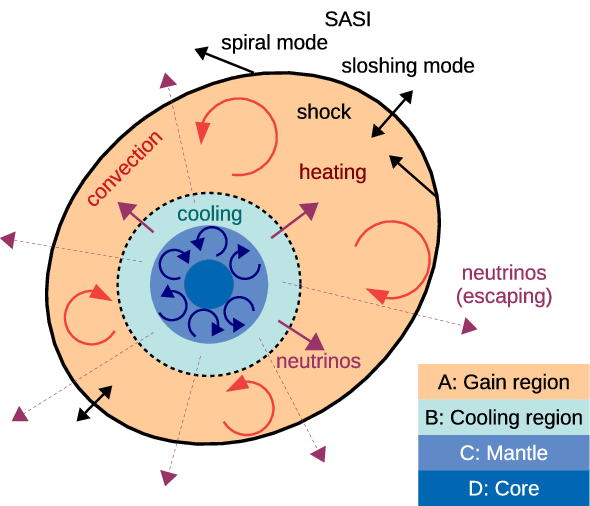}
    \caption{Structure of the inner region of a core-collapse supernova inside the shock (thick outer curve) during the post-bounce phase before shock revival. At the centre, there is a proto-neutron star consisting of the core and mantle. Neutrinos (bordeaux arrows) are emitted from the cooling region at the proto-neutron star surface. Most of the neutrinos escape, but some are reabsorbed in the heating region (orange) behind the shock. Heating in the gain region drives convective turnover motions. These support the development of explosions, e.g., by providing effective ``turbulent'' pressure
    to support shock expansion. Another hydrodynamic instability, the
    standing accretion shock instability (SASI), which leads to sloshing or spiral (rotational) motions of the shock, can play a similar supportive role. Cooling at the proto-neutron star surface also drives convection in its mantle.}
    \label{fig:mechanism}
\end{figure}

\subsection{Structure of the Supernova Core}
In the subsequent post-bounce phase, the stellar core consists of several distinct regions
(Figure~\ref{fig:mechanism}). At the centre, there is a proto-neutron star, which is still warm and relatively rich in protons with a neutron-to-proton ratio of about $3:1$. The proto-neutron star can be divided into a core ($\sim 0.5\ M_\odot$, region~D in Figure~\ref{fig:mechanism}) of low entropy and a continuously growing mantle (initially $\sim 0.9\ M_\odot$, region~C) of shock-heated material with higher-entropy. Below densities of a few $10^{13}\,\mathrm{g}\,\mathrm{cm}^{-3}$, neutrino cooling of accreted matter becomes very efficient, and the mantle transitions into a roughly isothermal atmosphere with a steep density gradient (region~B).
Outside this cooling region, a heating region (also called ``gain region'') develops between the proto-neutron star and the shock a few tens of milliseconds after bounce (region~A in Figure~\ref{fig:mechanism}). In this region, absorption of neutrinos from deeper layers outweighs neutrino emission, so that accreted material experiences net heating as it settles from the shock onto the proto-neutron star surface. The shock initially moves out only to a radius $100\texttt{-}200\,\mathrm{km}$, where it remains as a standing accretion shock.

For the dynamics of supernova explosions, multi-dimensional fluid flow in the supernova core plays a crucial role 
(\citealp{herant_94,burrows_95,janka_96}; for a review see
\citealt{Mueller2020}). In the gain region, convection can occur because heating by neutrinos results in an unstable entropy gradient. Similarly, neutrino cooling at the proto-neutron star surface drives convection in the mantle region. Under certain conditions, the gain region may also be subject to the standing accretion shock instability (SASI, \citealp{blondin_03,foglizzo_07,Oconnor2018}) due to a feedback cycle of acoustic and vorticity waves. Pre-collapse asymmetries in the progenitor star due to the presence of convection shells driven by nuclear burning will often jump-start the development of violent multi-dimensional flow behind the shock \citep{couch_15,mueller_17}. Furthermore,
turbulent fluid flow or differential rotation can drive
various dynamo amplification mechanisms that generate strong magnetic fields \citep{Akiyama2003,raynaud_20,matsuomoto_22,Mueller2024}.

\begin{figure}
    \centering
    \includegraphics[width=0.6\linewidth]{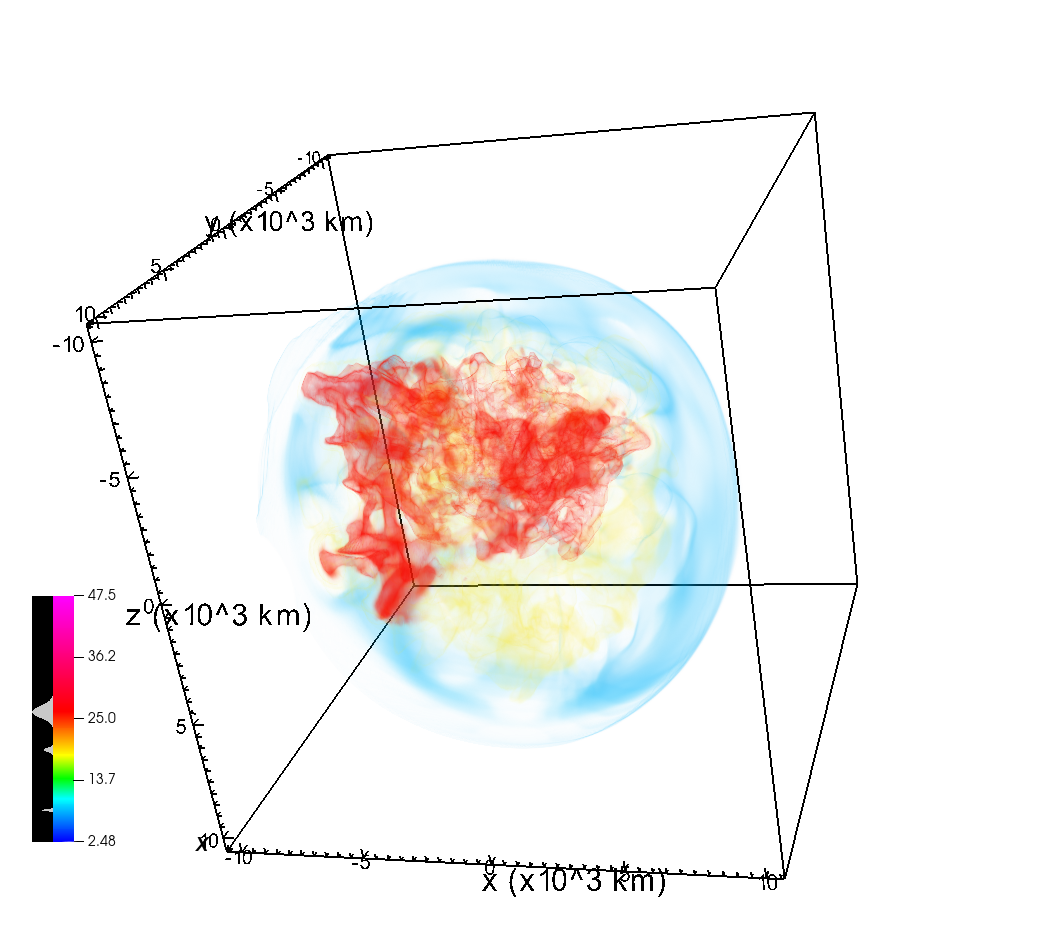}
    \caption{Three-dimensional simulation of the supernova explosion of a $3.5 M_\odot$ stripped-envelope progenitor 
    from \citet{Mueller2019},
    one second after the formation of a proto-neutron star and   several hundred milliseconds after the onset of the explosion. The volume rendering shows the entropy in the flow. The shock is visible as a blue translucent surface. It is pushed outward by expanding neutrino-heated bubbles (yellow, red). 
    The neutrino-driven outflows are strongly asymmetric, with considerably more neutrino-heated matter in one hemisphere. Much of the shocked matter is still channelled down to the proto-neutron star around the neutrino-heated bubbles. 
    The continuing accretion of material allows the proto-neutron star to maintain strong neutrino emission, which pumps more energy into the expanding bubbles.}
    \label{fig:expl3d}
\end{figure}

\subsection{Explosion Mechanisms}

Observations of supernovae and their progenitors indicate that massive stars with zero-age main sequence masses of probably up to $15\texttt{-}18\ M_\odot$ (for single stars) explode \citep{Smartt2015}. This implies that some mechanism eventually ``revives'' the stalled accretion shock and pumps enough energy into the post-shock matter to eject the outer shells and explain the kinetic energies of observed explosions, which range from
$\mathord{\sim} 10^{50}\,\mathrm{erg}$ to $\mathord{\sim} 2\times 10^{51}\,\mathrm{erg}$ for Type~IIP supernovae \citep[e.g.,][]{Pejcha2015} and may reach up to $\mathord{\sim} 10^{52}\,\mathrm{erg}$ for the rare class of broad-lined Ic supernovae, also known as \emph{hypernovae}, with very high ejecta velocities.
Several different explosion mechanisms have been proposed and studied extensively with numerical simulations.

In the \emph{neutrino-driven mechanism}, the explosion is provided by the energy deposition of neutrinos in the gain region \citep{mezzacappa_20}.
Over the first seconds of its existence, the proto-neutron star
radiates away several $10^{53}\,\mathrm{erg}$ in neutrinos.
The neutrino emission feeds on the internal energy of the proto-neutron star gained during the gravitational collapse. Even though only a small fraction of neutrinos are  reabsorbed in the gain region, neutrino heating can provide an energy input of about $10^{52}\,\mathrm{erg}$ over time scales of hundreds
of milliseconds to a few seconds according to detailed neutrino transport simulations. Much of this energy needs to be used up to unbind the heated material from the vicinity of the neutron star, however, leaving a residual energy in the range of
$\mathord{\sim} 10^{50} \texttt{-} 2\times 10^{51}\,\mathrm{erg}$
\citep{Mueller2017,Bollig2021,Burrows2024}, which is sufficient to explain the bulk of core-collapse supernova explosions. The detailed operation of the neutrino-driven mechanism is more complex than these simple order-of-magnitude considerations might insinuate, however. With the exception of the least massive core-collapse supernova progenitors, multi-dimensional effects in the supernova core are crucial for successful neutrino-driven explosions 
according to modern simulations. Multi-dimensional fluid flow
supports the development of neutrino-driven explosions by providing additional ``turbulent'' pressure and mixing of hot, neutrino-heated material in the gain region, which increase the radius of the accretion shock \citep[e.g.,][]{Mueller2020}. A sufficiently large shock radius is important for neutrino-driven shock revival, as this can help to increase the exposure time of accreted material to neutrino heating to the point where matter becomes unbound before settling on the proto-neutron star. Once the heating becomes sufficiently strong to exceed this threshold, it will trigger further shock expansion and a self-sustaining runaway process will set in \citep{Janka2001}.
Numerical simulations suggest that shock revival often occurs when the interface between the silicon and oxygen shell falls into the shock. Due to a density jump at this shell interface, its infall is associated with a drop in pre-shock ram pressure that is often sufficient to trigger such a runaway.

Importantly, the onset of the explosion will not stop further accretion onto the proto-neutron star immediately. 
Instead, matter will continue to be channeled from the shock to the proto-neutron stars in accretion downflows, while neutrino-heating drives away hot matter in other directions
(Figure~\ref{fig:expl3d}). 
This cycle of simultaneous accretion and mass ejections may last for several seconds and channel up to $\mathord{\sim} 0.1\,\mathrm{M}_\odot$ into 
neutrino-driven outflows.

Rare explosions with kinetic energies of more than 
$\mathord{\sim}2\times 10^{51}\,\mathrm{erg}$ cannot be readily explained by the neutrino-driven mechanism. The favoured explanation for such events is some form of \emph{magnetorotational mechanism} \citep{Akiyama2003,Burrows2007,Winteler2012,Obergaulinger2020,Aloy2021,Mueller2024}. In the classic form of the magnetorotational mechanism, magnetic fields extract the rotational energy $ E_\mathrm{rot}$ of a rapidly spinning proto-neutron star (ultimately coming from gravitational energy), which is roughly given in terms of its gravitational mass $M$, radius $R$, and spin period $P$ by
\begin{equation}
    E_\mathrm{rot}
    =2\times 10^{52}\,\mathrm{erg}
    \left(\frac{M}{1.4\  M_\odot}\right)
    \left(\frac{R}{12 \,\mathrm{km}}\right)^2
    \left(\frac{P}{1 \,\mathrm{ms}}\right)^{-2},
\end{equation}
and hence provides a sufficient reservoir to account for hypernova energies if extraction is efficient. Strong proto-neutron star surface fields of order $10^{15}\,\mathrm{G}$ are required
to extract the rotational energy on a short enough time scale to achieve shock revival. For the shortest possible neutron star spin periods of $\mathord{\sim}1\,\mathrm{ms}$
(to avoid centrifugal breakup), the maximum rotational energy roughly corresponds to the maximum kinetic energies inferred for hypernovae \citep{mazzali_14}. As an alternative to such (hypothetical) 
millisecond magnetars, the \emph{collapsar mechanism} relies on the rotational energy of a black hole accretion disk as a power source. The magnetohydrodynamically powered outflows can take the form of bipolar jets, but there may be a more isotropic, wind-like outflow component from the proto-neutron star or the disk as well. The polar regions around the compact object will eventually become strongly evacuated, and both the millisecond magnetar scenario and the collapsar scenario can accommodate the launching of relativistic jets as seen
in the gamma-ray bursts associated with many hypernovae \citep{Woosley2006}. There are two principal uncertainties about explosion scenarios involving rotation and magnetic fields. First, the problem of angular momentum transport in stellar interiors and hence the pre-collapse rotation as well as the magnetic field strengths and configurations in massive stars before core collapse are not yet fully understood (see Chapters ``Evolution and final fates of massive stars''). Second, the amplification of pre-collapse magnetic fields by compression during collapse is likely not sufficient to reach the field strengths required for magnetorotational explosion. Field amplification by magnetohydrodynamic instabilities after collapse must play a major role, but is challenging to study in simulations because of extreme requirements on spatial resolution \citep{moesta_15}. 

As an alternative to these two traditional scenarios, \emph{phase-transition driven explosions} have been proposed to occur in some massive stars. In this scenario, the proto-neutron star becomes unstable
when its core becomes dense enough to reach a putative first-order hadron-quark phase transition, collapses to a more compact state, rebounds, and thus launches a second shock wave that explodes the star \citep{Fischer2018}. Whether that phase transition is of first-order as required, and whether it could trigger explosions in a certain mass range are debated \citep{zha_21,Jakobus2022}.

\begin{figure}
    \centering
    \includegraphics[width=0.33\linewidth]{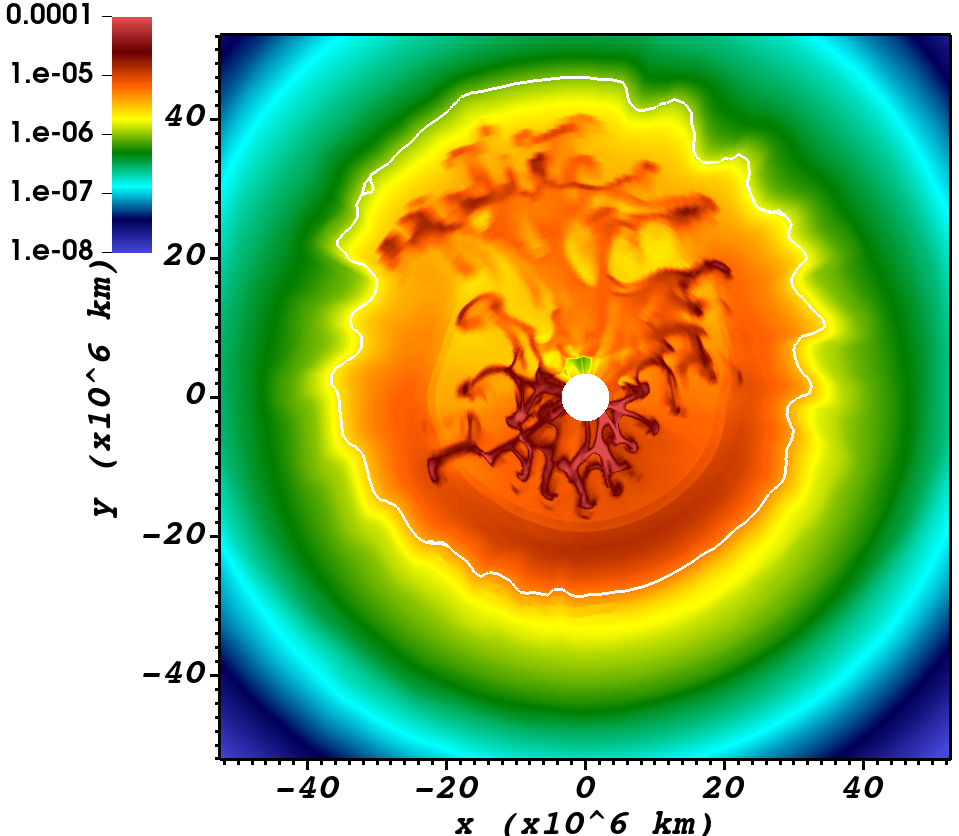}
    \includegraphics[width=0.33\linewidth]{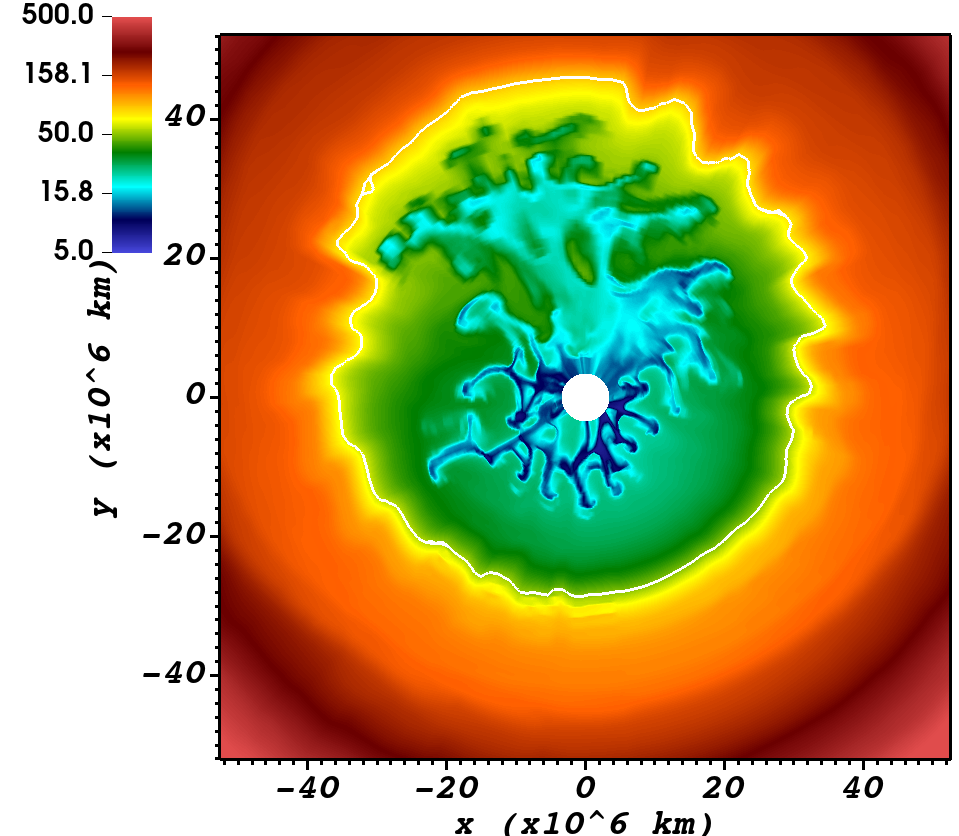}
    \includegraphics[width=0.33\linewidth]{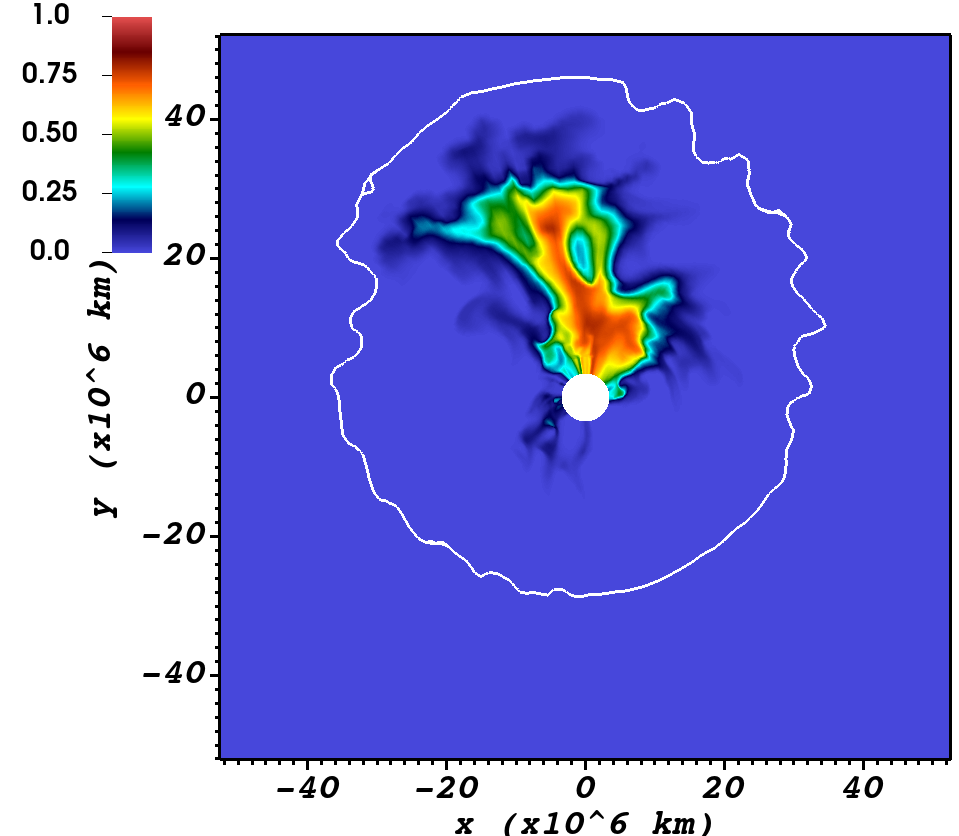}\\
    \includegraphics[width=0.33\linewidth]{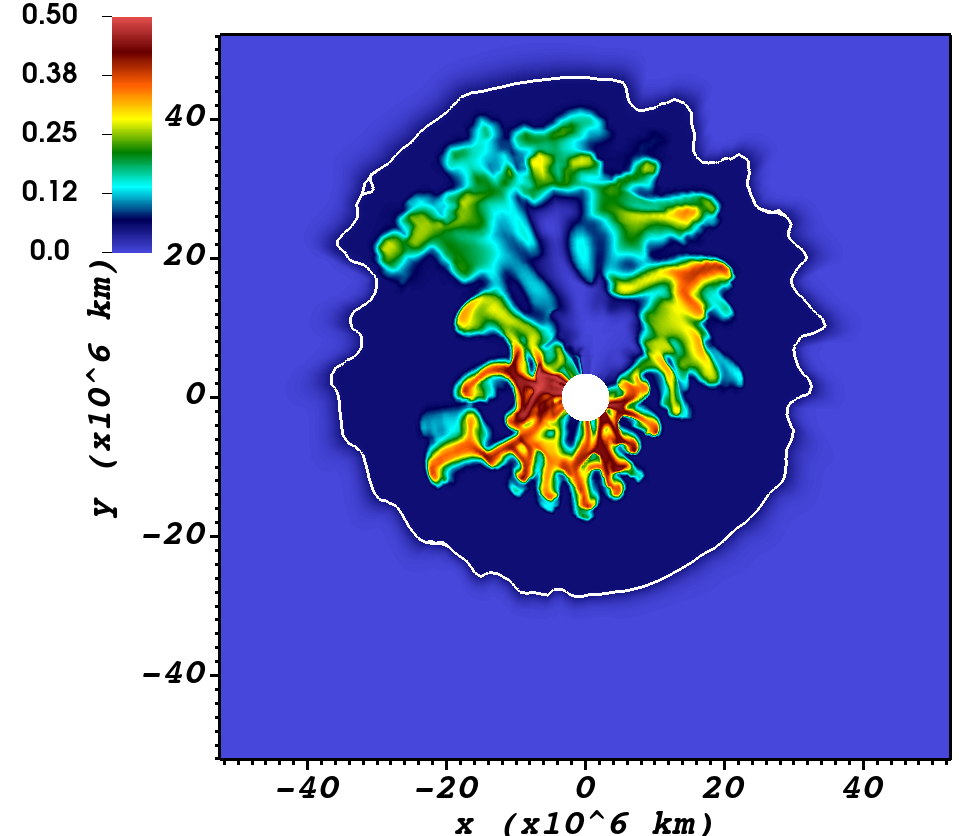}
    \includegraphics[width=0.33\linewidth]{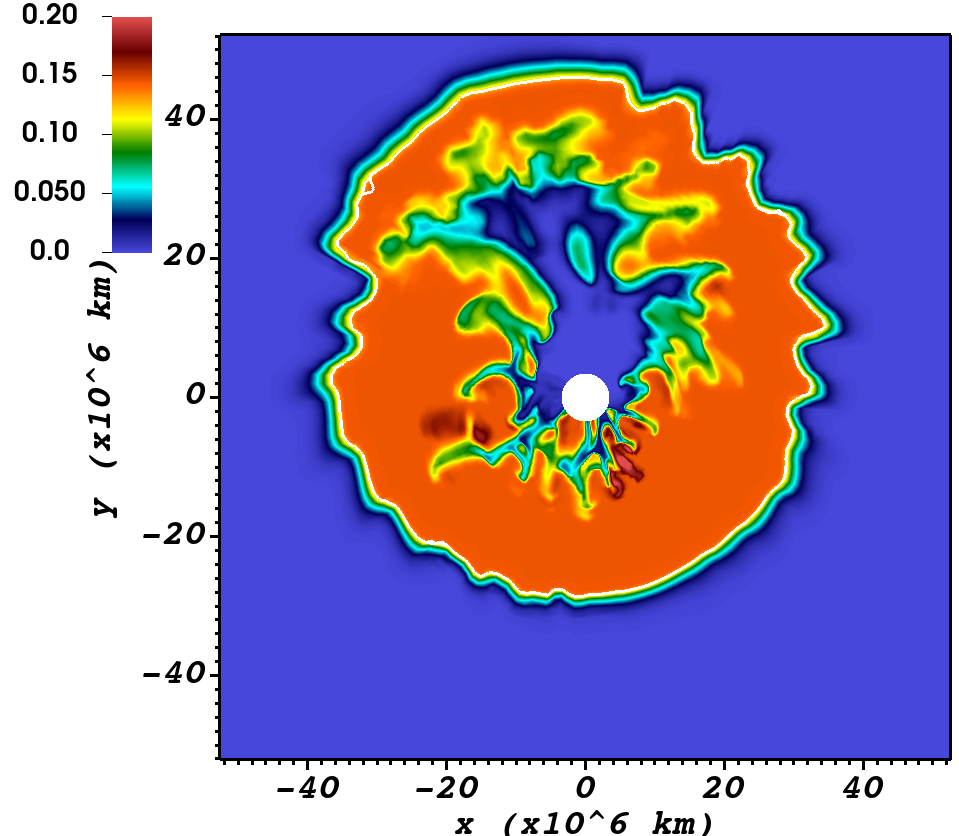}
    \includegraphics[width=0.33\linewidth]{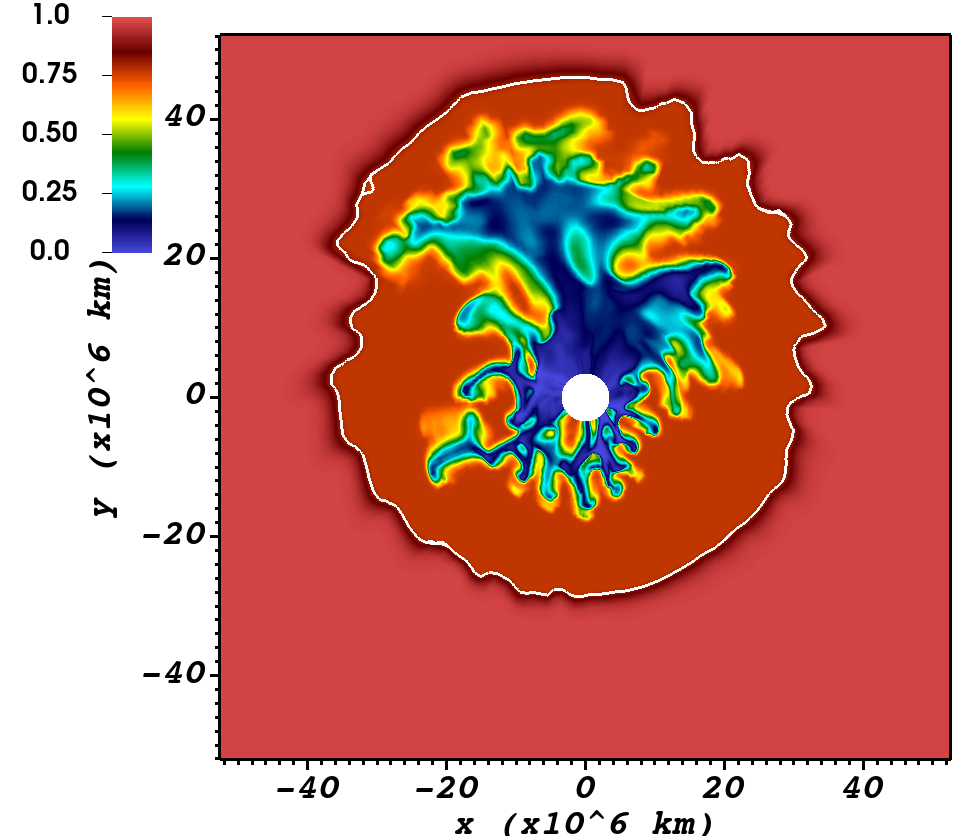}
    \caption{Rayleigh-Talyor instability in the simulation
    \citep{maunder_24}
    of a Type Ib/c supernova of a $3.5\  M_\odot$ stripped-envelope progenitor (same as in 
    Figure~\ref{fig:expl3d}. The figures show slices
    though the three-dimensional model about one hour after the onset of the explosion. The top row shows density (left, units of $\mathrm{g}\,\mathrm{cm}^{-3}$), entropy (middle, units of $k_\mathrm{B}$ per nucleon), and the mass fraction of iron-group elements (right). The bottom
    row shows the mass fractions of oxygen (left), carbon (middle) and helium (right). The
  white line indicates a helium mass fraction of $0.8$, and demarcates  
  the composition interface between the core and the unbured helium envelope, which has already been hit by the shock and expanded from its initial radius. Distinctive Rayleigh-Taylor plumes have formed and mix material from deep inside the star outward. Note that the large-scale
  asymmetry in the iron-group ejecta corresponds to the unipolar asymmetry of the early explosion in Figure~\ref{fig:expl3d}.}
    \label{fig:rayleigh_taylor}
\end{figure}

\subsection{Explosion Dynamics and Nucleosynthesis}
Both in the neutrino-driven scenario and in the magnetorotational
scenario, the ``engine'' continues to operate for significant time after shock revival, as we already alluded to in the previous section. The dynamics during this phase largely determines the properties of the explosion and the compact remnant. It also shapes the production of heavy elements (nucleosynthesis) during the explosion, which will influence the electromagnetic display of the supernova (Section~\ref{sec:lc_spec}), the composition and structure of supernova remnants (Section~\ref{sec:remnants}), and contributes to the chemical evolution of the host galaxy.

The dynamics of this early explosion phase, which lasts hundreds of milliseconds to seconds, has been explored most thoroughly in simulations of neutrino-driven explosions. The energy of such explosions is set by the amount of neutrino-heated matter that is marginally unbound from the vicinity of the proto-neutron star and then delivers an additional energy of $\mathord{\sim} 5\texttt{-}6 \,\mathrm{MeV}$ per nucleon due to the recombination of nucleons into $\alpha$-particles and then partially into heavy nuclei. The amount of ejected neutrino-heated material depends on how long the cycle of mass accretion and ejection is sustained, i.e., it is tightly coupled to the amount of matter accreted onto the proto-neutron star. Accretion eventually terminates when the explosion is strong enough for shocked matter to be accelerated roughly to the escape velocity to avoid being accreted \citep{marek_09,mueller_16}. Accretion will generally last longer in progenitors with massive and dense O shells around the Si-Fe core. After accretion has stopped, a neutrino-driven wind outflow from the neutron star will develop \citep{duncan_86}.

Since there are also outflows of neutrino-heated matter,
\emph{net} accretion onto the proto-neutron star is expected to terminate well before the explosion engine shuts off, often already a few hundred milliseconds after the shock is revived \citep{Mueller2019}. The dynamics of accretion is not only important for the final neutron star mass, however. Due to the asymmetries in the explosion, the neutron star can receive a substantial \emph{kick} to oppose the net momentum of the ejecta.
The acceleration of the neutron star occurs primarily by its gravitational attraction to the ejecta into the direction where the explosion is weaker (gravitational tug-bot mechanism, \citealt{scheck_06,Wongwathanarat_13}). Since the accretion downflows hit the neutron star with a finite impact parameter, they will deposit angular momentum and can spin up the neutron star even when the progenitor core had no rotation or slow rotation \citep{Wongwathanarat_13}.

The explosion dynamics also affects the composition of the inner ejecta. As the shock moves through the star and heats the shells around the core, it will trigger \emph{explosive burning} as long as the post-shock temperatures are high enough. Since the post-shock matter can be treated as a radiation-dominated gas, the
post-shock temperature $T_\mathrm{sh}$ is about
\begin{equation}
    T_\mathrm{sh}\approx
    \sqrt{\frac{3(\beta-1)}{a \beta} \rho v_\mathrm{sh}^2},
\end{equation}
in terms of the pre-shock density $\rho$, the shock velocity $v_\mathrm{sh}$ and the compression ratio $\beta \simeq 3\texttt{-}7$ in the shock; here $a$ is the radiation constant. 
About $3.5\times 10^9\,\mathrm{K}$ is required for explosive O burning into (mostly) Si and S, and about $5\times 10^9\,\mathrm{K}$ is needed to further burn Si and other intermediate-mass elements into iron-group elements \citep{iliadis_07}. Explosive burning to the iron-group will mostly produce ${}^{56}\mathrm{Ni}$ in shells that have no or at most very little neutron excess.
${}^{56}\mathrm{Ni}$ is also made by recombination of nucleons and $\alpha$-particles in the neutrino-heated matter that is ejected from the vicinity of the proto-neutron star;
it will later play a crucial role for the CCSN light curve. However, the nucleosynthesis becomes more complicated if the shock-heated matter expands rapidly. Under these conditions, freeze-out of hot ejecta in nuclear statistical equilibrium  will leave an excess abundance of $\alpha$-particles whose reactions on heavier nuclei modify the abundances seen in ``normal'' freeze-out. This $\alpha$-rich freeze-out is the main source for $^{44}\mathrm{Ti}$, whose decay is an important power source
for the later phase of the light curve (Section~\ref{sec:lc_spec})
and the emission from supernova remnant (Section~\ref{sec:remnants}). For significant neutron- or proton-excess, more complex nucleosynthesis processes can occur 
\citep[see, e.g.,][]{Woosley2002,wanajo_23}
and produce, e.g., elements beyond the iron-group \citep{wanajo_11}.

\subsection{Evolution to Shock Breakout}
After one of the aforementioned mechanisms has powered up the explosion, the shock wave still needs to propagate to the stellar surface until the explosion becomes visible by electromagnetic radiation. This can take hours to days in red supergiant stars with extended envelopes, and as a little as several minutes in very compact stripped-envelope progenitors.

As the shock propagates to the surface, the shock velocity $v_\mathrm{sh}$ changes as the kinetic energy of the explosion is distributed over a large amount of material, and as the pre-shock density declines. A rough approximation for $v_\mathrm{sh}$
in terms of the explosion energy $E_\mathrm{exp}$,
ejecta mass $M_\mathrm{ej}$, pre-shock density
$\rho$ and shock radius is given by \citep{Matzner1999}
\begin{equation}
\label{eq:vshock}
v_\mathrm{sh}(r)
\approx 
0.794
\left(\frac{E_\mathrm{exp}}{M_\mathrm{ej}}\right)^{1/2}
\left(\frac{M_\mathrm{ej}}{\rho(r) r^3}\right)^{0.19}.
\end{equation}
Although the shock may transiently accelerate at shell interfaces in the star where the density drops strongly, the ejecta \emph{behind} the shock generally experience deceleration by a positive pressure gradient (bigger pressure outside) behind the shock. Such a decelerated flow is subject to the \emph{Rayleigh-Taylor instability} \citep{Chevalier1976,Mueller1991,Ye2017} if denser, low-entropy material comes to lie below lighter, high-entropy material; in some sense the flow behaves as under an outward-directed ``effective gravity''. The Rayleigh-Taylor instability can drive significant outward mixing of heavy elements from the inner shells of the explosion, and drag light elements downwards
(Figure~\ref{fig:rayleigh_taylor}). Other fluid instabilities such as the Richtmyer-Meshkov instability \citep{Ye2017} may also occur. 

These mixing instabilities are seeded by the initial asymmetries that inherently develop during the first few second of the explosion in the neutrino-driven and magnetorotational scenario. How much mixing occurs as the shock propagates through the envelope depends both on the initial seed asymmetries as well as the progenitor structure \citep{Mueller2020}. Red supergiant stars tend to experience significant mixing by the Rayleigh-Taylor instability with significant development of small-scale plumes. Explosions of compact stripped-envelope progenitors tend to experience less mixing so that the ejecta geometry more strongly reflects the initial explosion asymmetries.

Generally, the dynamics of the ejecta behind the shock can be rather complex. In addition to mixing instabilities, episodes of acceleration and deceleration of the shock can lead to the formation of \emph{reverse shocks}, which can both trigger and inhibit mixing, and decelerate material to a degree that it will fall back onto the compact remnant.

Such \emph{fallback} is thought to play a most prominent role in massive progenitors with big helium and carbon-oxygen cores. In such progenitors, 
accretion may continue for a long time after the shock is revived, and eventually push the neutron star over its maximum mass towards black hole collapse. Furthermore, because of a large binding energy of the envelope, the explosion energy may be insufficient to eject all the matter outside the proto-neutron star. The dynamics of fallback may be of significant importance for shaping the mass distribution of black holes. For compact objects, a quantitative comparison with the observed populations of neutron stars and black holes is intertwined with the problem of single and binary stellar evolution.
One important insight from recent years is that neutrino-driven explosions produce a wide range of gravitational neutron star masses, reaching to $1.7 \ M_\odot$ and beyond \citep{Muller2017,Burrows2024}. This is in line with observations \citep{Oezel2016} that have overturned the long-held assumption that neutron stars are generically born with masses around $1.35\ M_\odot$.

Due to adiabatic cooling of the ejecta, the dynamics of the expanding blast waves become less affected by pressure effects as the shock moves closer to and eventually breaks out of the stellar surface. Beyond shock breakout, large parts of the ejecta therefore approach homologous expansion, i.e., the radial velocity becomes proportional to radius. Homologous expansion only holds approximately, however. Complex dynamics involving reverse shocks and fallback can continue after shock breakout, and the interaction with the circumstellar environment is a major factor during the supernova remnant phase
(Section~\ref{sec:remnants}) and in some instances for the dynamics of the
observable transient (Section~\ref{sec:lc_spec}). Radioactive decay and strong outflows from the compact remnant will affect both the thermal evolution and hence the electromagnetic emission from the ejecta, as well as their kinematic and spatial structure.
For example, heating by the decay of radioactive $^{56}\mathrm{Ni}$ can further accelerate
and inflate Ni-rich bubbles \citep{blondin_01,gabler_21}. Repercussions of this 
``Ni bubble effect'' and outflows such as pulsar winds can be studied directly both from nebular spectra (Section~\ref{sec:lc_spec}) and in the remnant phase (Section~\ref{sec:remnants}).

\section{Light curves and spectra}
\label{sec:lc_spec}

\begin{figure}
\centering
\includegraphics[width=0.8\linewidth]{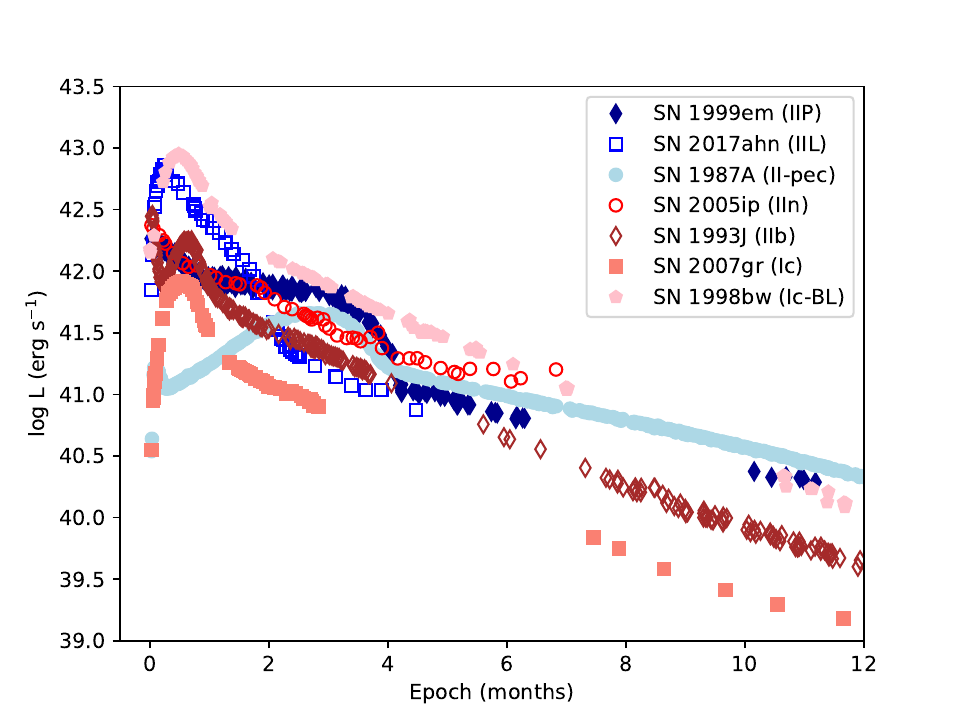}
\caption{Pseudo-bolometric ($UBVRI$, no correction for light outside these bands) light curves of characteristic SNe from each of the main classes (no Type Ib SN is plotted but these are similar to Type IIb and Ic SNe). Data from \citet{Menzies1987,Catchpole1987,Barbon1995,Richmond1996,Patat2001,Sollerman2002,Elmhamdi2003,Galama1998,Valenti2008,Hunter2009,Stritzinger2012,Tartaglia2021}.}
\label{fig:LCs}
\end{figure}

As the shock breaks out of the star, the electromagnetic display begins. 
Figure \ref{fig:LCs} shows a compilation of light curves for the different SN types discussed in section \ref{chap1:sec1}. We see that SNe reach peak brightness sometime between 10-100d, with a characteristic luminosity value of $\sim 10^{42}$ erg s$^{-1}$. Variations around that luminosity is  about a factor 10 up and down ($10^{41}-10^{43}$ erg s$^{-1}$). After a relatively steep decline from peak (end of "diffusion phase"), the light curves show a kink as the SN goes into the "tail" phase. The brightness declines over time to a characteristic value $\sim 10^{40}$ erg s$^{-1}$ by 1 year. The different phases, and the driving mechanisms and key parameters for each, will be described below.

\subsection{Shock breakout phase}
\label{sec:breakout}
\begin{minipage}{0.49\linewidth}
The electromagnetic display begins when the shock wave reaches the stellar surface, or, if the star has a dense wind, as the shock approaches the optically thin point in the wind. The actual shock breakout gives a very brief transient (minutes/hours), with most radiation in X-rays and UV. Some of this hard radiation will photoionize the CSM which then emits recombination lines over a few hours or days \citep{Quimby2007}.  As the star gives no sign of its impeding collapse, observation of the shock breakout phase can occur only serendipitously, which has happened only a few times, e.g. for SN 2008D \citep{Soderberg2008,Modjaz2009shock} and SN 2016gkg \citep{Bersten2018}. Shock breakout observations hold promise to diagnose in particular progenitor radius (larger radii give more energetic bursts), shock velocity (higher velocity gives higher-frequency spectral energy distribution), and very late mass loss (impacts the time evolution); see \citet{Waxman2017} for a theory review. 
One of the topics on the current research fore-front of shock breakout is to understand how the intrinsic 3D structure of the stellar envelope affects the shock break-out signal \citep[e.g.][Fig. \ref{fig:goldberg}]{Goldberg2022}.
\end{minipage}
\begin{minipage}{0.45\linewidth}
\centering
\includegraphics[width=0.75\linewidth]{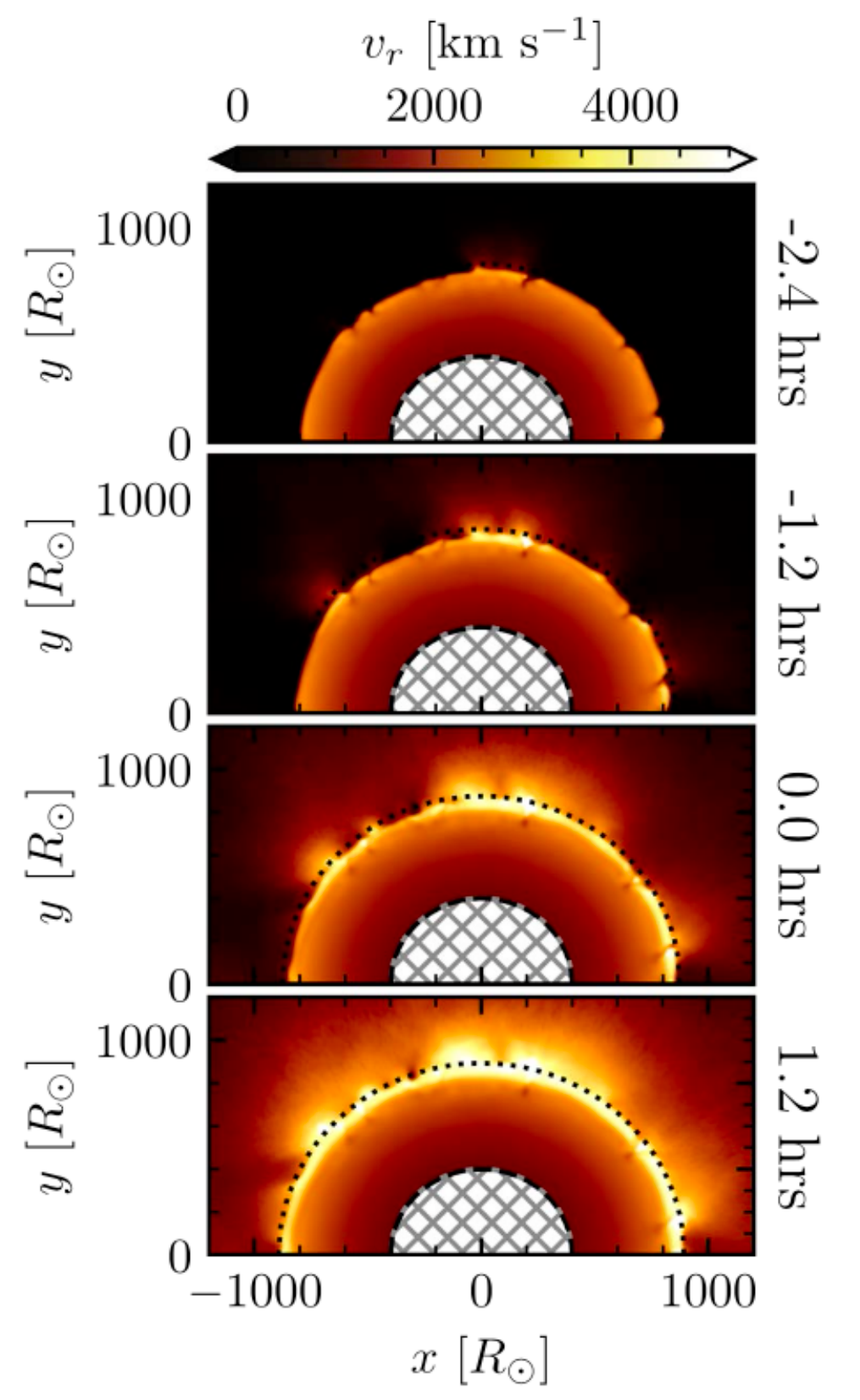}
\captionof{figure}{Shock breakout simulated in 3D for a RSG. From \citet{Goldberg2022}.}
\label{fig:goldberg}
\end{minipage}

\subsection{Diffusion phase}
\label{sec:diffphase}
\blue{Following shock breakout, the light curve enters the \emph{diffusion phase}, governed by the diffusive release of internal energy on a time-scale of weeks to months.
The early light curve evolves in a way that depends on how much internal energy has been deposited by the shock passage, what fraction of this energy will be radiated rather than adiabatically converted to bulk kinetic expansion energy, and on the time scale of the diffusion process. By equipartition, the initial internal energy is close to half of the total shock-deposited energy, the value of which is given by the outcome of the competing inward ram pressure and outward thermal  pressure described in Section \ref{sec:bernhard}. The radiation efficiency depends mainly on the progenitor radius $R_0$  \citep[scaling approximately with $\sim R_0^{0.8}$; luminosity $L \propto E^{5/6} M_{ej}^{-1/2} R_0^{2/3} \kappa^{-1/3}$, diffusion time $\Delta t \propto E^{-1/6} M_{ej}^{1/2} R_0^{1/6} \kappa^{1/6}$ $\rightarrow E_{rad}/E = L \times \Delta t/E \propto E^{-1/3} R_0^{5/6}$,][where $\kappa$ is a space and time averaged opacity]{Popov1993,Kasen2009},
and is for all progenitors $\ll 1$; i.e. almost all energy deposited by the shock will be converted to kinetic bulk expansion energy of the stellar debris rather than being radiated away. For a Red Supergiant ($R_0 \approx 500\ R_\odot$), its value is around 1\%, giving $\sim 10^{51}\times 0.01 = 10^{49}$ erg of emitted radiation. 
The time scale of diffusive energy release $\Delta t$ depends most sensitively on the ejecta mass, scaling with $M_{\rm ej}^{1/2}$,
(or $M_{\rm ej}^{3/4}$, depending on the role of recombination). 
For RSGs $M_{\rm ej} \approx 10$ $M_\odot$ which gives a time-scale of a few months. In these Type IIP SNe the bolometric light curve is initially declining (although the optical light curves rise as cooling brings an increasing fraction of the spectral energy distribution from X-ray/UV into the optical), then a recombination wave sets in after about three weeks, as the ejecta temperature reaches the H recombination temperature of $\sim$6000 K, which flattens the light curve to a plateau. The brightness on this plateau depends most sensitively on
 the explosion energy ($L \sim E^{5/6}$) and observations give values of $0.1-1$ B (1 B = $10^{51}$ erg) \citep[e.g.][]{Martinez2022}.}

\blue{Radioactive elements produced in the explosion also influence the diffusion phase light-curve. In Type IIP SNe the radioactivity extends the plateau by about 20\% by sustaining and boosting the opacity \citep[e.g.][]{Kasen2009}, while in compact star  ($\lesssim\ 50\ R_\odot$) explosions, where the explosion-deposited internal energy becomes  strongly degraded by adiabatic expansion, it plays a yet larger role. For explosions of Blue Supergiants (BSG) and He cores, the internal energy provided by the radioactive decay chain $^{56}$Ni$\rightarrow^{56}$Co$\rightarrow^{56}$Fe governs the light curve from a few days already.
The powering mechanism is mainly that gamma-rays emitted in the radioactive decay Compton scatter in the ejecta, a process in which high-energy electrons are created which subsequently heat, ionize and excite the gas. The total energy provided by radioactivity is not overly large; $E_{\rm decay} = 2 \times 10^{49} \times \left(M(^{56}\mbox{Ni})/0.1\ M_\odot\right)$ erg, but the delayed injection (decay time scales 8.8d and 111d for the two steps) avoids the strong adiabatic degradation that the shock-deposited energy suffers from. For radioactivity-powered light curves (Type Ib, Ic, IIb), the diffusion phase brightness and duration give information about the mass of $^{56}$Ni (approximately proportional to the peak luminosity) and the quantity $\kappa^{1/2} M_{\rm ej}^{3/4} 
 E^{-1/4}$ (approximately proportional to the diffusion phase duration for SESNe, note difference to expression for Type IIP SNe above). Ejecta mass is more robustly inferred than explosion energy due to the stronger scaling ($M_{\rm ej}^{3/4}$ vs $E^{-1/4}$). Typical inferred values are $M(^{56}\mbox{Ni}) \sim 0.1~M_\odot$ and $M_{\rm ej} = \left(1-5\right) M_\odot \times E_{51}^{1/3} \kappa_{0.1}^{-2/3}$, with the energy in units of $10^{51}$ erg and opacity in units of 0.1 cm$^2$g$^{-1}$ \citep[e.g.][]{Drout2011,Bersten2012,Prentice2016}. With detailed modelling \citep[e.g.][]{Blinnikov1998} and combining with spectral information, degeneracies with $E$ and $\kappa$ can be mitigated - for example is the photospheric velocity around peak light, inferable e.g. from the O I 7774 absorption line, a good indicator of the quantity $\sqrt{2E/M_{\rm ej}}$ \citep{Dessart2015}. Detailed light curve modelling has also shown that symmetry breaking is strong in the explosion, with good fits requiring the initially innermost layers (rich in $^{56}$Ni) to be strongly mixed outwards over the minutes and hours after explosion \citep[e.g.][]{Shigeyama1990}; this provides important constraints on the explosion hydrodynamics and the various scenarios and mechanisms discussed in Section 2.  
 }

\blue{In addition to explosion-deposited energy and radioactivity, another power source of potential importance is circumstellar interaction (CSI). A steady wind from the progenitor star sets up a $\rho(r) \propto r^{-2}$ circumstellar medium (CSM), containing a mass of 
\begin{equation}
M_{\rm CSM}(r) = 3 \times 10^{-3}\ M_\odot \left(\dot{M}/10^{-5} M_\odot \mbox{yr}^{-1}\right) \left(v_w/10\ \mbox{km s}^{-1}\right)^{-1} \left(r/10^{16}\ \mbox{cm}\right),
\end{equation}
where $\dot{M}$ is the mass-loss rate and $v_w$ is the wind velocity. While this may seem small, if only a percent or so of the SN kinetic energy ($\sim$$10^{51}$ erg) is converted to radiation by collision with this CSM, this could overtake the radiated energy budget. Should $\dot{M}/v_w$ be much larger than the canonical RSG wind value ($\sim 6 \times 10^{14}$ g cm$^{-1}$), this will be the case as the swept-up CSM mass within the diffusion time then becomes $\gtrsim$ 1\% of the SN ejecta mass. Observations of very early light curves and spectra have indicated that some massive stars also appear to eject large amounts of mass in eruptive episodes prior to explosion \citep{Ofek2013}, which is not yet well theoretically understood. Such late mass loss will further increase the role of CSI \citep{Margutti2017}.} 

A final power source that may come into play is the compact remnant; if a fast-rotating, highly magnetized neutron star is formed, but spin-down occurs over weeks or months rather than seconds as needed in the magnetorotational explosion scenario (Section 2), a pulsar wind is formed that may provide most of the energy input to the nebula. We see such a situation today in the Crab nebula (Section \ref{sec:remnants}), and it is also a contender to explain the light curves of some unusually bright SNe over weeks and months \citep{Maeda2007,Woosley2010,Inserra2013,Nicholl2013,Lunnan2018,Moriya2018}.

\textbf{Spectra.}
\blue{In the diffusion phase the expanding nebula is optically thick, and spectral formation can be reasonably well described as blackbody radiation emitted from a photosphere experiencing scattering in the outermost layers. The density profile in the outer line-forming region is quite steep, a $v^{-7}$ power law or yet steeper \citep{Matzner1999}. The rapid, differential expansion leads to a characteristic line profile called a \emph{P-Cygni} line, where absorption is seen on the blue side of the rest wavelength and emission on the red side - the peak is at the rest wavelength which facilitates line identifications. In Type II SNe the photospheric spectra probe largely unprocessed, H-rich gas \citep[although dredge-up of CNO-burning products can occur in some progenitors,][]{Fransson1989} and is therefore a good metallicity probe \citep{Dessart2013,Anderson2018}, whereas in SESNe also the outermost layers contain nuclear processed material, from H and/or He burning. For this latter class, signatures of He, O, Na, Ca, Si and Fe are typically seen in photospheric spectra \citep[e.g.][]{Liu2016,Shahbandeh2022}.
The time evolution of photospheric and line scattering velocities, inferred from the P-Cygni profiles, give important information on the SN envelope structure. The line transfer in supernovae (or in any other nebulae with rapid differential expansion) behave in a particular way; photons scatter in the resonance layer in the same way for any $\tau_S \gtrsim 1$, where $\tau_S$ is the so-called Sobolev optical depth, a local quantity. P-Cygni lines therefore give upper (no absorption) or lower (absorption) limits to ion densities rather than absolute values, which makes abundance determinations more challenging than in static nebulae. Because the photosphere recedes towards lower velocities with time, more structure in the spectrum emerges continuously, which reduces line blending and facilitates the composition analysis.}

\textbf{Polarimetry.} Information about the geometry of the SN may be obtained by photometric polarimetry, or, for bright enough sources,  spectroscopic polarimetry \citep[see][for a review]{Wang2008}. Asymmetry in the free electron morphology induces polarization in Thomson-scattered light, and asymmetry in element distributions induces asymmetry in the P-Cygni lines. Observations give typical polarization levels varying between $\sim$0.1\% up to a few percent. Close-to axisymmetric morphologies (e.g. from some magneto-rotational explosion scenarios) would give a straight line in the so-called $Q/U$ plane, whereas 3D morphologies with small clumps would give broad, scattered distributions. The most commonly observed pattern - a loop - instead indicates 3D morphologies with quite large structures \citep[plume sizes $R_{cl} \gtrsim 1/4 R_{phot}$, ][]{Tanaka2017}. Such a morphology for the outer SN material is broadly consistent with what is seen in the spatially resolved Cas A remnant (see section \ref{sec:CasA}). For the deeper lying regions, a good moment for spectropolarimetry is when the SN is transitioning from the photospheric to the nebular phase - the large asymmetries of the core then combine with still lingering opacity to give a maximum in the polarization signal \citep{Leonard2006}. While the data clearly show significant asymmetries, there is also much diversity \citep{Nagao2024}. While some constraints have been derived based on crafted multi-D ejecta \citep{Dessart2024}, more detailed conclusions await the computation of polarization signals from 3D explosion simulations and comparison with data.  

\subsection{Tail phase}
\blue{The expanding nebula eventually (after 1-4 months, depending on the ejecta mass, explosion energy and powering mechanism) reaches an extended and dilute enough state that generated radiation can escape on short time-scale with little or no interaction. The SN is then said to have entered the \emph{tail phase} or the \emph{nebular phase} \citep[see][for a review]{Jerkstrand2017}.}

\blue{\textbf{Light curves.} The bolometric light curve in this phase follows the instantaneous radioactive energy deposition, which equals the radioactive decay rate times a trapping factor, which is unity for full trapping but can be below unity if decay particles like gamma rays partially escape. The effective opacity for gamma rays is about 0.03 cm$^2$g$^{-1}$, and the trapping time is of order (expression derived from setting the radial gamma-ray optical depth to unity for a uniform sphere):}
\begin{equation}
t_{\gamma-trap}=100\mbox{d} \left(\frac{M_{\rm ej}}{3\ M_\odot}\right)\left(\frac{E}{10^{51}\ \mbox{erg}}\right)^{-1/2}.
\label{eq:ttrap}
\end{equation}
\blue{The early tail phase in Type IIP SNe ($t \sim 120-200$d, $M_{\rm ej} \sim 10\ M_\odot$) has close to full trapping and therefore offers a good opportunity to determine the amount of $^{56}$Ni synthesised - values $0.01-0.1\ M_\odot$ are typically found \citep{Rodriguez2021}. Mapping out how the $^{56}$Ni yield depends on the $M_{\rm ej}$ and $E$ parameters, inferrable from the diffusion phase \citep[e.g.][]{Muller2017}, puts important constraints on the SN explosion mechanism. 
If the bolometric luminosity can be determined for epochs later than 2-3 yr, the masses of other radioisotopes such as $^{57}$Ni and $^{44}$Ti can also be estimated, as these take over the powering at late epochs. In particular, $^{44}$Ti is a sensitive probe of the innermost, high-entropy regions originating from just outside the proto-NS \citep{Wongwathanarat2017}. Direct detection of the gamma-ray/X-ray lines emitted in the $^{44}$Ti decay both for SN 1987A \citep{Boggs2015} and Cas A \citep{Renaud2006} have here been of value.
} 

\blue{In SESNe the $M_{\rm ej} E^{-1/2}$ quantity is typically a factor 2-10 lower than in Type IIP SNe (Sec. \ref{sec:diffphase}), and Eq. \ref{eq:ttrap} shows that gamma-ray trapping is typically incomplete already in the early tail phase ($\sim$50-100d for these SNe), so $^{56}$Ni mass estimates from the tail are for these SNe model-dependent. One can fit $M(^{56}\mbox{Ni})$ and $t_{\gamma-trap}$ simultaneously, the results give a median $^{56}$Ni mass of 0.08 $M_\odot$ and a 
 median trapping time of 117d \citep{Afsariardchi2021}. The $^{56}$Ni mass values inferred from the tail are in good agreement with those inferred from the diffusion phase brightness, and the mean value is a factor 2-3 higher than in Type IIP SNe, a difference driven by an apparent dearth of SESNe with $^{56}$Ni masses below $\sim$0.03 $M_\odot$. Population synthesis studies considering binary mass loss for SESN progenitors have provided explanations for this difference in $^{56}$Ni masses between the SN classes  \citep{Schneider2021}.} 

\blue{\textbf{Spectra.}
\begin{figure}
\centering
\includegraphics[width=1\linewidth]{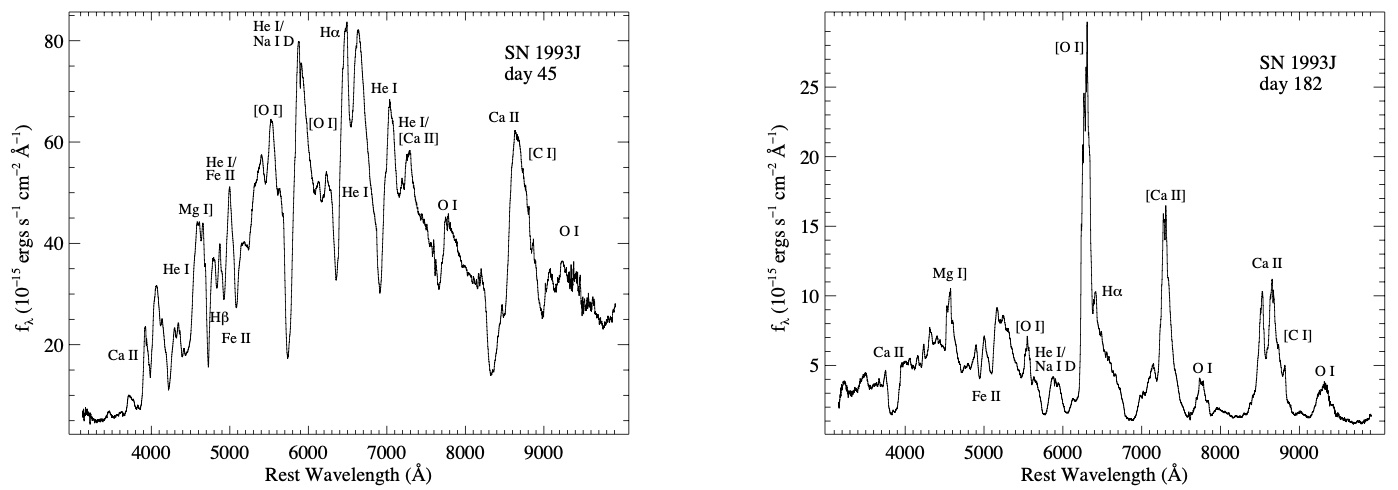}
\caption{Spectral evolution as the SN transitions from the photospheric phase (left) to the nebular phase (right), here for the Type IIb SN 1993J. In the photopheric phase the spectrum shows P-Cygni profiles with blueshifted absorption and redshifted emission (but with peak at the rest wavelength). In the nebular phase the spectrum takes on an emission-line character. From \citet{Matheson2000-1}.}
\label{fig:matheson2000}
\end{figure}
The spectrum in the tail phase changes character to an emission-line spectrum (Fig. \ref{fig:matheson2000}, right panel). The widths of the emission lines reflect the kinematics, with expansion velocities (thousands of km/s) being much larger than the thermal velocities (few km/s). The dominant emission lines come from low-lying metastable states in neutral and singly ionized atoms of abundant elements. These can be populated by  thermal collisional excitations at the characteristic nebula temperature (a few thousand degrees); examples are [O I] 6300,~6364, [Fe II] 7155, and [Ca II] 7291,~7323\footnote{[] means a forbidden transition (occurs via magnetic dipole (M1) or electric quadropole (E2) transition), while ] means semi-forbidden (occurs via electric dipole (E1) transition with spin change).}, which provide much of the cooling  \citep{Kozma1998-I}. For some lines, for example H$\alpha$, Mg I] 4571 and O I 9263, population of the parent state by recombination can also be important. By combining information from several lines one may attempt deduction of physical parameters and composition for the inner, metal-rich core of the SN, which is only visible in this phase. This allows the direct testing of modern theories for massive star and SN  nucleosynthesis \citep[e.g.][]{Heger2003,Chieffi2013}. The densities are in the nebular phase too low for a Local Thermodynamic Equilibrium (LTE) assumption in general, and modelling requires solving non-LTE (NLTE) rate equations. Because the stellar nucleosynthesis production of oxygen is a monotonic and strongly increasing function of stellar mass (Fig. \ref{fig:oxmass}), inference of the oxygen mass from nebular lines may be used to constrain the progenitor star mass. The fact that oxygen has several strong and relatively unblended lines in the optical from its dominant neutral ionization stage, enables this to be done in a relatively robust way. Oxygen masses in the range 0.1-1 $M_\odot$ are inferred for the vast majority of CCSNe, in line with expected production in $M_{ZAMS}\lesssim 18-20\ M_\odot$ stars \citep{Jerkstrand2014,Jerkstrand2015a,Jerkstrand2018,Silverman2017,Dessart2021}. Higher yields, implicating more massive stars, appear mostly associated with Ic-BL SNe \citep{Sollerman2000,Mazzali2001,Jerkstrand2017SLSN}, a subclass of the Type Ic category with unusually high velocities, often associated with GRBs. However, a few cases of massive progenitors giving regular Type Ibc SNe also exist \citep[e.g.][]{Valenti2012,Emir2023}. These events are too rare to account for all or most stars over $M_{ZAMS} \approx 20\ M_\odot$, which seems to suggest that most massive progenitors fail to explode, but some succeed by, presumably, the operation by the magneto-rotational or collapsar mechanisms discussed in Section \ref{sec:bernhard}. Signatures of other hydrostatically made elements include He \citep{Li1995}, N \citep{Barmentloo2024} and Mg \citep{Jerkstrand2015a}.  Information on explosive nucleosynthesis can be obtained from lines such as [Fe II] 7155 and [Ni II] 7378, the latter which probes changes in the neutron-to-proton ratio in stellar evolution and/or explosion \citep{Fröhlich2006,Maeda2007-2006aj,Jerkstrand2015b,Jerkstrand2015c}. The mid-infrared iron-group lines are in turn good diagnostics of the $^{56}$Ni-bubble expansion discussed in Section 1, having shown that the relatively small mass of radioactive material expands to fill $\gtrsim$ 20\% of the SN core volume  \citep{Li1993,Jerkstrand2012}.}

\begin{figure}
\includegraphics[width=0.49\linewidth]{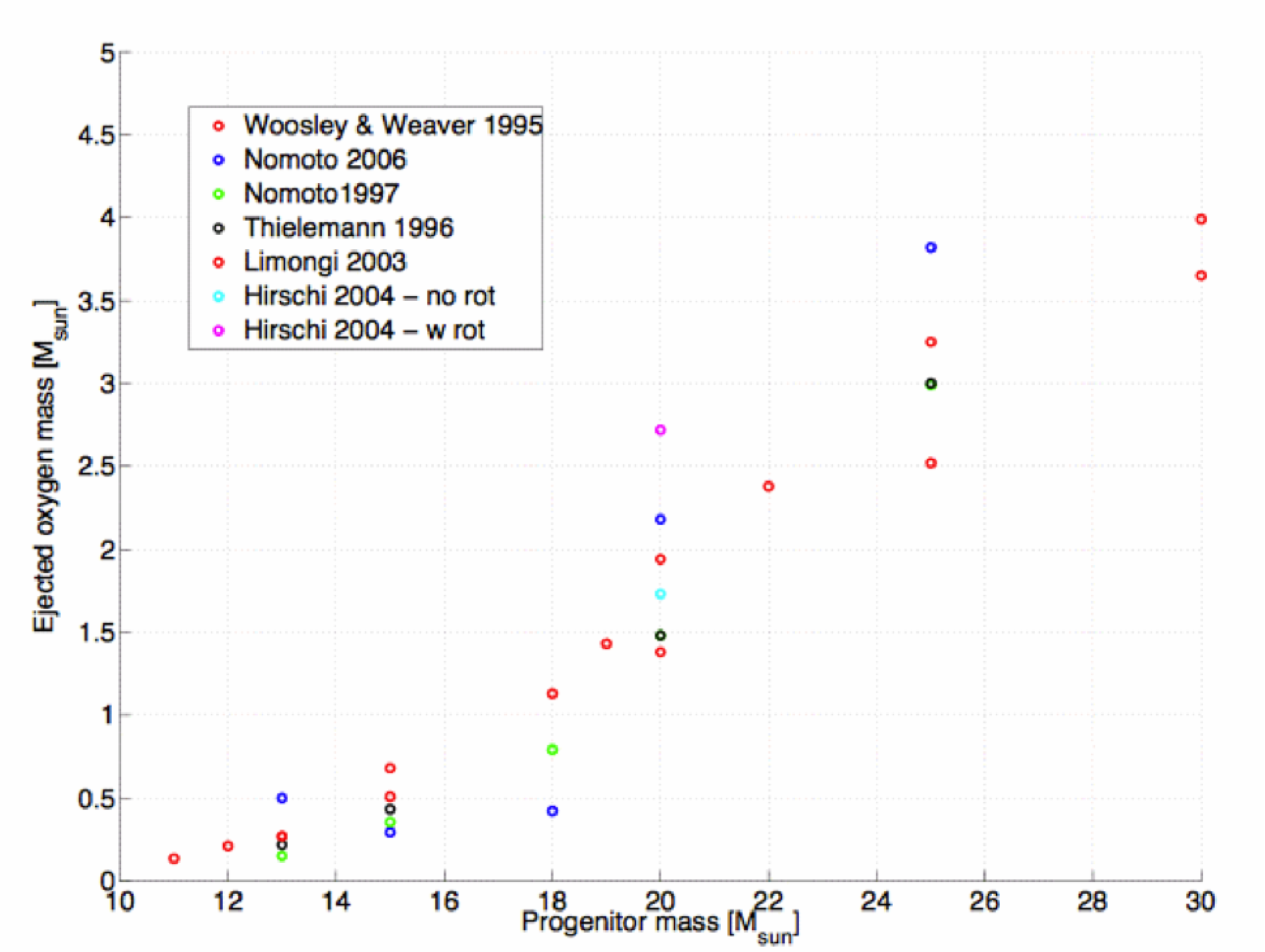}
\includegraphics[width=0.49\linewidth]{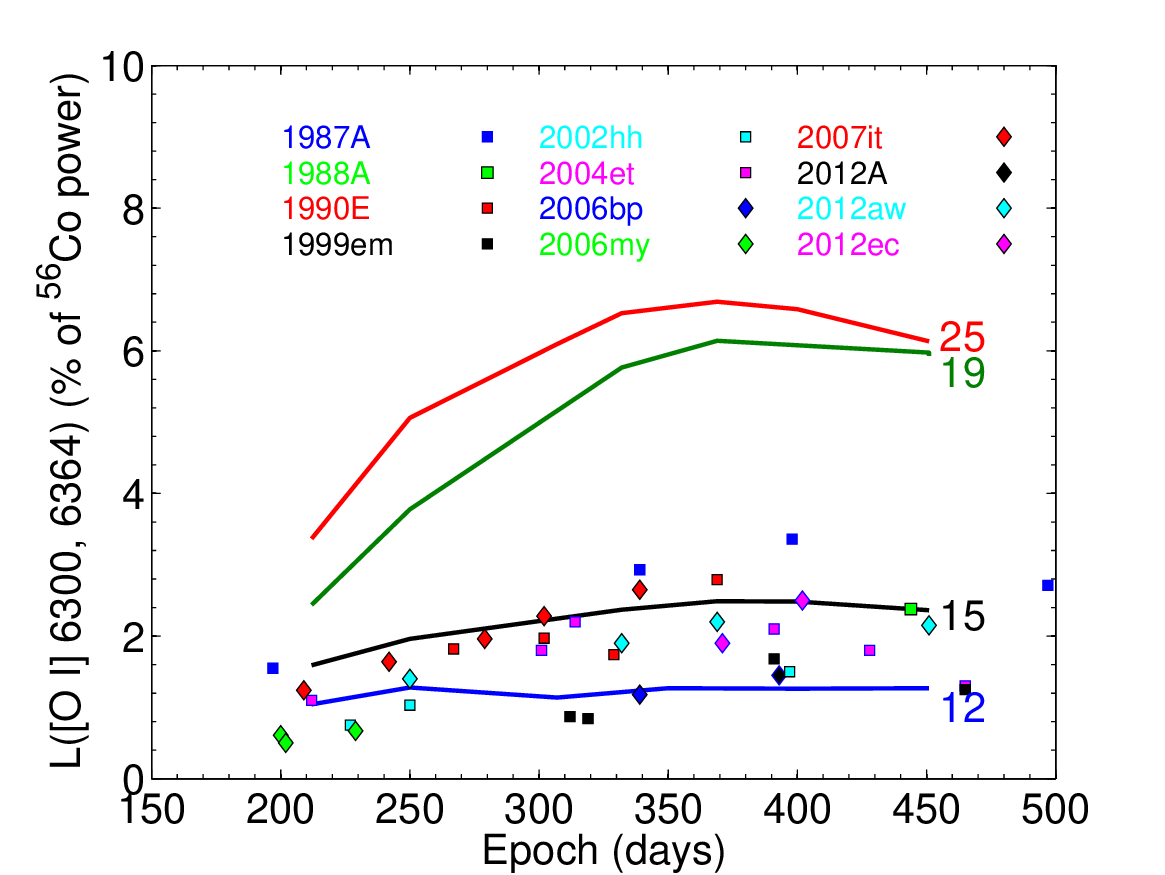}
\caption{\emph{Left:} Theoretical SN yields of oxygen as function of progenitor mass ($M_{ZAMS}$). Data from \citet{Woosley1995,Thielemann1996,Nomoto1997,Nomoto2006,Limongi2003,Hirschi2004}. \emph{Right:} Luminosities in [O I] 6300, 6364 for twelwe Type IIP SNe compared to models for different $M_{ZAMS} (12,15,19,25\ M_\odot)$. From \citet{Jerkstrand2015b}.}
\label{fig:oxmass}
\end{figure}

\blue{It has long been known that nebular lines profiles provide a rich source of information on the multi-dimensional structure of the nebula \citep{Filippenko1989,Spyromilio1993}. In homology, ejecta material in sheets perpendicular to the line-of-sight (LOS) has the same LOS velocity, and for a given spectral line, each wavelength shift $\Delta \lambda$ relative to line center therefore probes emissivity by the sheet moving with corresponding line-of-sight velocity $v_{LOS} = \Delta \lambda/\lambda_0 c$. By this tomography effect, bulk line properties (width, shift, skewness) probe the large-scale morphology, whereas fine-structure (wiggles, spikes) probes clumping and fragmentation. Observed line profiles indicate emissivity over a broad range of velocities for most elements, indicating strong hydrodynamic mixing, in line with results from light curve modelling. The mixing has been inferred to occur on macroscopic but not microscopic scales, so different elements still probe different, compositionally distinct, nuclear burning layers in the star. Fine-structure in line profiles is observed to be quite constant over time, and similar between different lines of the same ion \citep{Matheson2000-2}, which indicates that variations in physical conditions over time and space do not swamp out the underlying morphology of an element distribution. The complex debris field obtained in modern neutrino-driven explosion simulations give a wide diversity of line profiles \citep{vanBaal2023}, including the often observed double-peaked ones \citep{Mazzali2005,Modjaz2008,Maeda2008,Taubenberger2009,Mili2010,Fang2022}. By comparing the domain spanned by all viewing angles to observations, specific explosion models can be tested \citep{vanBaal2024}.}

\blue{The tail phase also sees the activation of a rich chemistry in CCSNe, with initially ($t \gtrsim$ few months) molecules such as CO and SiO forming in the inner, metal-rich regions, and later ($t \gtrsim$ years) dust \citep[see][for a review]{Sarangi2018}. Dust forms in the cooling expanding gas of SNe just as it does around cool stars, in planetary nebulae and novae. These processes can alter the physical conditions in the nebula, allow new probes through IR and radio emission by molecules and dust, and make SNe important cosmic dust factories. The significance of the cosmic dust contribution depends on how efficiently the formed dust is destroyed again when strong shock waves are later formed as the SN enters the supernova remnant phase. The composition of SN dust is an active area of research, with possibilities  including silicates \citep{Rho2008}, amorphous carbon \citep{Matsuura2015}, and metal needles \citep{Dwek2004}.}

\section{Supernova remnant phase}
\label{sec:remnants}

As the SN continues to homologously expand in its nebular phase, it continuously sweeps up more and more of the CSM and eventually reaches the interstellar medium (ISM). The timescales for when the CSM and ISM are encountered by the blast wave depend on the forward shock velocity, as well as properties of the mass loss.
The characteristic velocity of the supernova ejecta $v_{\rm ej}$ having a typical kinetic energy $E_{\rm SN} \sim 10^{51}\, {\rm erg} \sim (1/2)M_{\rm ej}v^{2}_{\rm ej}$ can be expressed as 

\begin{equation}
    v_{\rm ej} \sim 10^{4}\ {\rm km\ s}^{-1} \left( \frac{E_{\rm SN}}{10^{51} \rm erg} \right)^{1/2} \left( \frac{M_{\rm ej}}{M_{\odot}} \right)^{-1/2},
\end{equation}

\noindent where $M_{\rm ej}$ is the mass of the SN ejecta. The forward shock will decelerate as it sweeps up material, while a reverse shock develops that moves opposite in direction of the forward shock (in the Lagrangian frame) and excites the slower moving ejecta \citep{Raymond2018}. Between the shocked supernova ejecta and shocked CSM/ISM, which each have different composition and entropy, is an area of pressure equilibrium called the contact discontinuity. The shock waves of SNRs can also accelerate particles to relativistic energies and are a source of cosmic rays \citep{Blasi13}.

\begin{figure}[tp]
\centering
\includegraphics[width=0.5\linewidth]{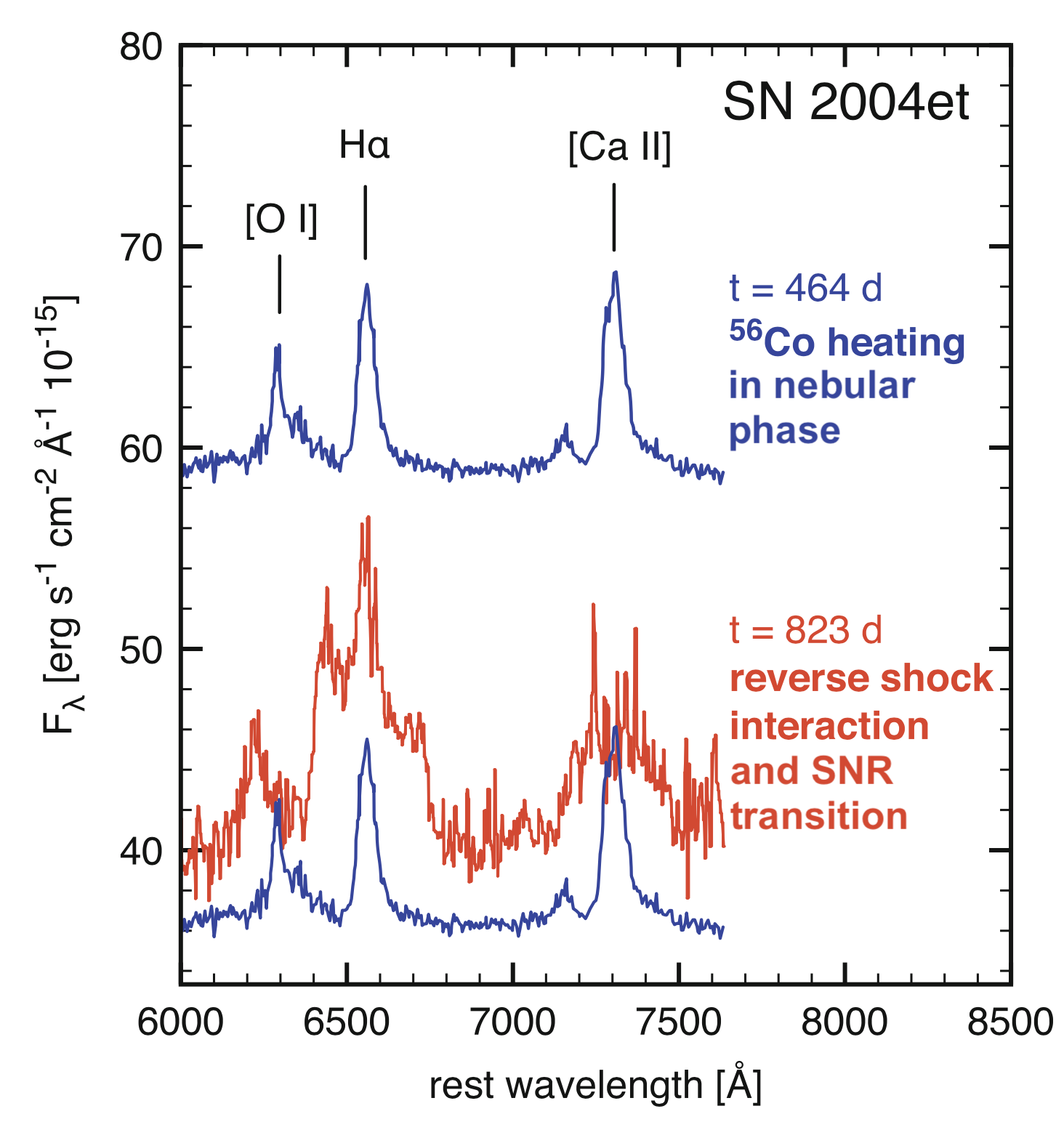}

\caption{Optical spectra of SN 2004et showing its transition from the nebular to supernova remnant phase. At t=464 d, the emission from the [O I], H$\alpha$, and [Ca II] lines is powered by radioactive $^{56}$Co. At t=823 d, the kinematic distribution of these ions undergo a dramatic broadening, consistent with emission now being dominated by the reverse shock exciting outer, higher velocity ejecta. Asymmetry in the profile is due to internal extinction of optically-emitting material from the far side of the supernova by dust. Note that the t=823 d spectrum will likely also have new contributions from [N II] 6548, 6583 and [O II] 7319, 7330. Adapted from \citet{MF17} using data from \citet{Kotak2009}.}
\vspace*{-3mm}
\label{fig:sn-to-snr}
\end{figure}

The lifespan of an SNR is approximately $10^{5-6}$\,yr, and its evolution is often grouped into four stages (see, e.g., \citealt{Bamba2022}). In the first stage of \textit{free expansion}, the stellar ejecta retain their initial velocity without significant deceleration and expand homologously such that the radius of the blast wave is $r(t)=v_{\rm ej}t$, where $t$ is time since explosion. The blast wave gains mass as it sweeps up surrounding material of density $\rho(r)$, where $\rho(r)_{\rm CSM} \propto r^{-2}$ for wind-like mass loss and $\rho(r)_{\rm ISM} \propto \rho_{\circ}$ for constant density ISM. The free expansion stage ends when the mass swept up $M_{\rm swept} \sim M_{\rm ej}$. If expansion is primarily in an ISM-like environment, this transition occurs at a radius of


\begin{equation}
    r_{\rm free\ expansion} \sim 2\,{\rm pc} \left( \frac{M_{\rm ej}}{M_{\odot}} \right)^{1/3}\left( \frac{\rho_{\circ}}{10^{-24}\rm \,g\,cm^{-3}} \right)^{-1/3},
\end{equation}

\noindent and time 

\begin{equation}
    t_{\rm free\ expansion} \sim \frac{r}{v_{\rm ej}} \sim 200\,\rm yr\left( \frac{E_{\rm SN}}{10^{51} \rm erg} \right)^{-1/2} \left( \frac{M_{\rm ej}}{M_{\odot}} \right)^{5/6}\left( \frac{\rho_{\circ}}{10^{-24}\rm \,g\,cm^{-3}} \right)^{-1/3}.
\end{equation}

Subsequent remnant phases include the adiabatic \textit{Sedov} phase ($t \sim 10^{2-4}$ yr; $r \sim t^{2/5}$; M$_{\rm ej}$ $<$ M$_{\rm swept}$) and the pressure-driven \textit{snowplow} phase ($t \sim 10^{4-5}$ yr; $r \sim t^{1/4}$; M$_{\rm ej}$ $\ll$ M$_{\rm swept}$) that develops as the SN interacts with more and more of the surrounding material, the gas cools radiatively at constant momentum, and the SN retains only vestiges of the initial ejecta mass and its distribution. In the final \textit{dissipation} phase, the shock slows down to a velocity comparable to the sound velocity, and the SNR merges with the ISM.  

Throughout its evolution, the SNR will sweep up a substantial amount of ISM ($\sim 10^{4}$\,M$_{\odot}$), and the interaction results in multi-wavelength emission. Radio emission is largely from synchrotron radiation associated with electrons accelerated in regions run over by the forward shock, whereas the ionized broad inner ejecta region that has been reverse-shocked emits across infrared-through-X-ray wavelengths as a result of various thermal and non-thermal processes. The forward-shocked CSM/ISM can also emit at these wavelengths. After timescales of $10^{5-6}$\,yr, the ejecta expand into the dilute plasma of the ISM, filling the space between stars, and create cavities that can reach up to several hundred parsecs in diameter. Molecules and dust in the ejecta feed into molecular clouds, which in turn become the building blocks for new stars and planetary systems. Overall SNRs are vital contributors to the heating and chemical evolution of galaxies.  

There is no generally accepted definition for the point when a SN is said to have become a SNR. Historically, the SNR phase has been attributed to the time when a SN departs from free expansion and begins to strongly interact with its surrounding environment, or when both UV/optical line and continuum emission from a SN's ejecta fall below that generated by interaction with either surrounding CSM/ISM or via emission from a central compact stellar remnant \citep{Fesen2001}. However, such generalizations fail to incorporate scenarios when SN-CSM interaction is strong immediately after explosion, as in the case of SNe IIn. Sometimes the transition from supernova to supernova remnant is defined simply when it reaches an age of $\gtrsim 100$ years \citep{BW2017}. 

A convenient definition, applicable for most cases when there is no strong SN-CSM interaction immediately after the SN, is the time at which SN–CSM interaction leading to reverse shock heating of inner ejecta dominates observed emission \citep{MF17}. At these epochs, which follows from the late tail phase, broad, boxy lines of H$\alpha$ and/or metal lines such as He I 5876 and  [O III] 4959, 5007 \citep[e.g.][]{Matheson2000-1} are observed in optical spectra. These broad lines ($v \gtrsim 4000$ km\,s$^{-1}$), which can often be observed simultaneously with narrower lines associated with forward shocked CSM ($v \lesssim 2000$ km\,s$^{-1}$), arise from the fast, outermost layers of the SN, absorbing X-rays from the reverse shock created by the CSM collision \citep{Chevalier1994}. Often asymmetry is observed in profiles of these line emissions, giving a false sense of preferential blueshifted ejecta velocity, that is in fact attributable to dust within the ejecta. Conspicuous line substructure is a common and long-lasting phenomenon in the late-time spectra, and is linked to large scale coherent structure in the ejecta \citep{Mili2010}. For some core-collapse H-rich SNe, the timescale to transition to the SNR stage as defined this way can be as short 1–2 years (see, e.g., \citealt{Black2017,Dessart2023}; Figure~\ref{fig:sn-to-snr}). 

Whereas most supernova remnants in the Milky Way were initially identified as extended radio sources with nonthermal radio spectra, the majority of extragalactic supernova remnants have been found via optical narrow-band imaging surveys that utilize the line strength diagnostic ratio [S II]/H$\alpha$ $>$ 0.4 \citep{Long17}. Most extragalactic SNRs are unresolved or barely resolved ($r_{\rm SNR} \lesssim 1^{\prime\prime}$), and lack an associated classification because they are evolved ($>1000$ yr) and the original supernova was not observed. The possibility to monitor extragalactic supernovae many years after discovery was first recognized in the late 1980s with the optical re-detections of SN 1980K \citep{FB88} and SN 1957D \citep{Long1989}. Since then, there have been several examples of continued monitoring of SNe as they transition between supernova phases (diffusion and tail) to the supernova remnant phase \citep{Milisavljevic2012}. In the majority of cases when late-time monitoring is possible, some long-lived energy source related to interaction between the SN and ISM/CSM, or contributions from a pulsar/magnetar star, is present. In the case of SN 1987A in the LMC, a fortuitous nearby distance has made it possible to  monitor the supernova continuously from detection of neutrinos to transition to the supernova remnant phase \citep{McCray2017}. 

\begin{figure}[tp]
\centering
\includegraphics[width=0.9\linewidth]{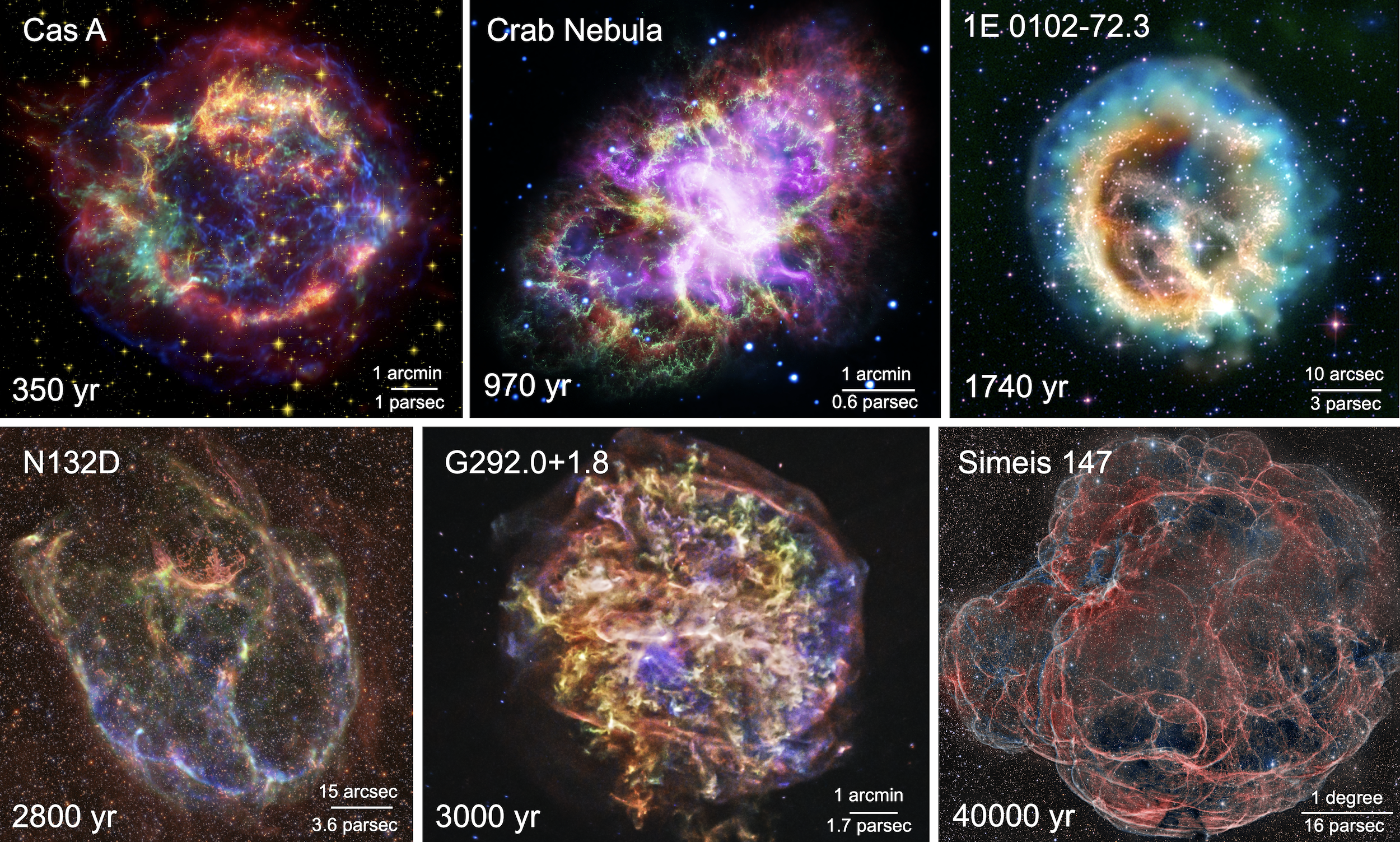}

\caption{Supernova remnants of core collapse supernovae. Approximate ages and physical scales adopted from \citet{Reed95}, \citet{Hester2008}, \citet{Banovetz2023}, \citet{Temim2022}, and  \citet{Ng2007}. Image credits: Cas A (NASA/CXC/SAO/STScI/JPL- Caltech/Steward/O. Krause et al.), Crab Nebula (NASA, ESA, and J. DePasquale, STScI), E0102 (NASA/SAO/CXC/STScI/MIT/D.\ Dewey et al./J.\ DePasquale), N132D (NASA/CXC/CXO/HST/STScI/J.\ Schmidt), G292 (NASA/CXC/SAO), and Simeis 147 (Georges Attard).}
\vspace*{-3mm}
\label{fig:snrs}
\end{figure}

\subsection{Galactic Supernova Remnants}

Currently, there are approximately 300 SNRs catalogued in the Milky Way \citep{Green2019}.\footnote{An up-to-date listing can be found at \url{https://www.mrao.cam.ac.uk/surveys/snrs/}. See also \url{http://snrcat.physics.umanitoba.ca/}.} Multiple diagnostics exist to distinguish whether remnants are associated with core-collapse versus Type Ia supernovae. A fundamental means is to look for the presence of a compact object (neutron star or black hole) or to perform spectroscopy of a light echo of the original supernova (if present). Various methods at X-ray wavelengths exist, including estimates of iron abundance \citep{Reynolds2007}, X-ray line morphologies \citep{Lopez2009}, and Fe--K line energy centroids \citep{Patnaude2015}. Attempts at relating specific SN types with SNRs can also involve connecting SNRs with historical records of ``guest stars'' as has been possible with the Crab Nebula \citep{CS77}.\footnote{The term ``guest star'' is a literal translation from ancient Chinese astronomical records and refers to a star that suddenly appears in the sky where no star had been observed before, and then fades away after some time. Modern astronomy recognizes that these ``guest stars'' were actually manifestations of cataclysmic events such as novae and supernovae.} However, since these ancient sightings lack spectroscopic information, and are often incomplete in terms of the precise location of the SN and the evolution of the light curve, they can lead to uncertain conclusions. The Crab is the only Milky Way SNR that has a firm historical account of the parent supernova associated with a core collapse explosion.

The morphology of a supernova remnant is dependent on both the explosion dynamics and the environment into which the supernova expands into. Figure~\ref{fig:snrs} shows core collapse supernova remnants with a variety of morphologies reflecting diverse progenitor systems, environments, and ages. Most young ($\lesssim 1000$ yr) galactic SNRs have strong individual characteristics that are not always shared by other young SNRs or easily connected to SN subclasses. Differences largely arise from the role CSM/ISM can play in affecting remnant's properties and evolution. The surrounding environment will have been sculpted by prior mass loss of the progenitor system; the mass loss may be a smooth wind or clumpy or eruptive; and the gas may deviate from a homogeneous distribution due to effects of rotation or the influence of a binary companion. Asymmetries intrinsic to the SN explosion can also contribute to the morphology, especially for SNRs still in free expansion. The evolution of massive stars toward the ends of their life cycles is likely to be nonspherical and may have extensive inter-shell mixing \citep{AM11}, and hence asymmetries introduced by a turbulent progenitor star interior can seed explosion asymmetries that persist until the SNR phase. Large-scale magnetic fields can also contribute to SNR morphology at advanced ages \citep{West2016}.

SNRs are typically divided into (i) shell-like remnants, in which emission comes almost entirely from a distinct shell, (ii) filled-center or Crab-like remnants, which have a central pulsar wind nebula (PWN) and whose brightness decreases radially outward, and (iii) composite remnants that blend shell and Crab-like properties. Shell-like remnants pass through the typical four stages of supernova remnant evolution, whereas the dynamics of Crab-like remnants are more complex because the central source provides electromagnetic radiation and high-energy particles that influence surrounding gas and the SNR evolution. Approximately 80\% of SNRs are shell-like, 10\% are composite, and a few percent are Crab-like. In addition to these classifications is a relatively rare number of \textit{mixed morphology} remnants that exhibit a shell-like morphology in the radio band and centrally peaked thermal
emission in the X-ray band \citep{Rho98}. Unlike composite remnants, mixed morphology remnants lack a PWN. 

SNRs where hydrogen emission is weak or non-existent in the ejecta and is instead dominated by heavy metals including oxygen, neon, sulfur, argon, and calcium, are known as oxygen-rich SNRs. O-rich SNRs likely originate from massive stars that were largely stripped of their hydrogen envelopes. Only a handful of O-rich remnants have so far been identified in our Galaxy and in nearby galaxies. Cassiopeia A (Cas A), which is regarded as the prototypical O-rich SNR \citep{Fesen01}, Puppis A \citep{WK85}, and G292+1.8 \citep{Goss1977} are among the Galactic members. N132D \citep{DD1976} and 0540-69.3 \citep{Mathewson1980} are in the Large Magellanic Cloud. 1E 0102.2-7219 \citep{Dopita1980}, 0103-72.6 \citep{Park2003}, and B0049-73.6 \citep{Hendrick2005} are in the Small Magellanic Cloud. Several mixed-morphology SNRs, including CTB 1, HB 3, and W28, have also shown evidence of O-rich ejecta \citep{LS2006,Pannuti2010,Pannuti2017}. An ultraluminous O-rich SNR has been found in NGC 4449 \citep{BH1978,MF2008}.

\begin{figure}[tp]
\centering
\includegraphics[width=0.98\linewidth]{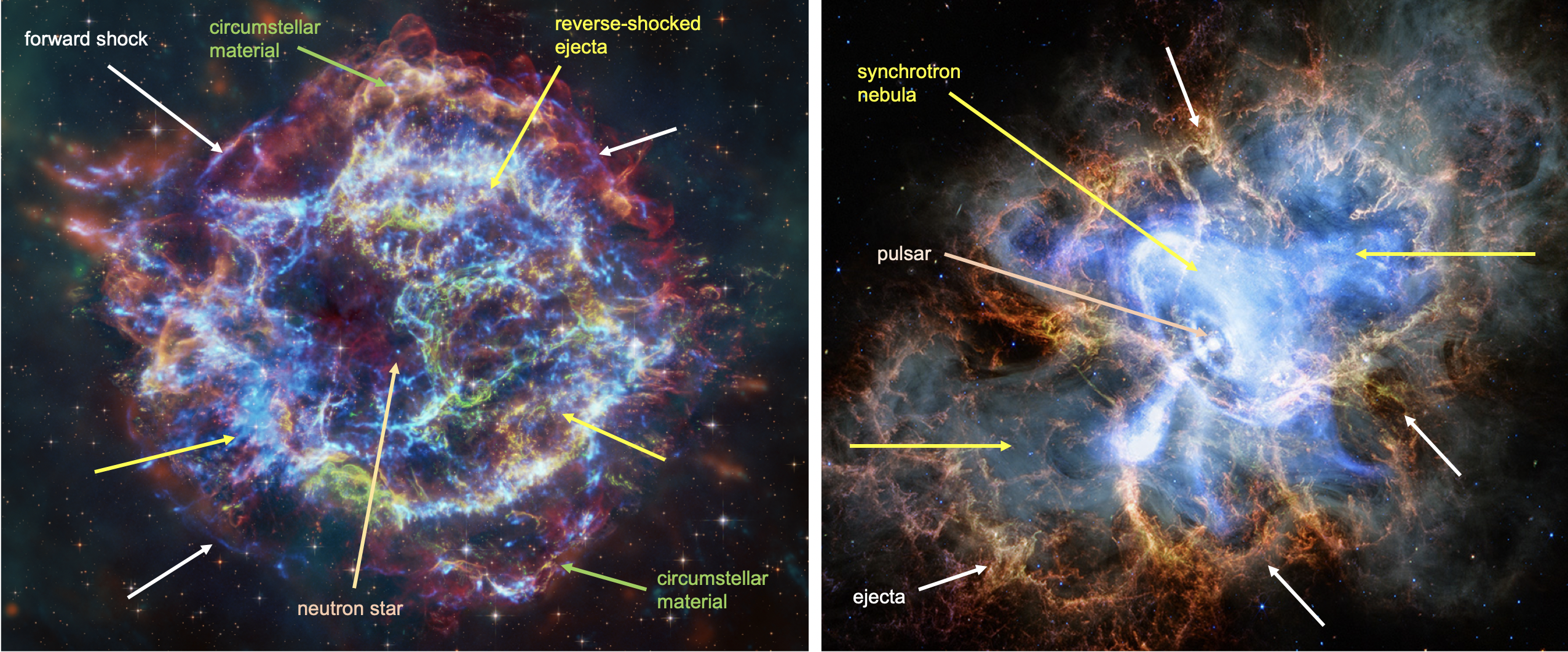}

\caption{Multiwavelength images of Cassiopeia A [X-ray: NASA/CXC/SAO; Optical: NASA/ESA/STScl; IR: NASA/ESA/CSA/STScl/Milisavljevic et al., NASA/JPL/CalTech; Image Processing: NASA/CXC/SAO/J.\ Schmidt and K.\ Arcand] and the Crab Nebula [X-ray, Chandra: NASA/CXC/SAO; Infrared, Webb: NASA/STScI/T.\ Temim et al.; Image Processing: NASA/CXC/SAO/J.\ Major].}
\vspace*{-3mm}
\label{fig:casa-crab}
\end{figure}

\subsection{Cassiopeia A - The prototypical core-collapse SNR}
\label{sec:CasA}

Cas~A (Figure~\ref{fig:casa-crab}) is the youngest Galactic core-collapse SNR known ($\approx 350$ yr; \citealt{Fesen06}). This youth and a relatively nearby distance ($3.4^{+0.3}_{-0.1}$ kpc; \citealt{Reed95,Alarie14}) have made Cas A extremely important to investigate the physics of core collapse through resolved, multiwavelength observations. Although there is no credible historical record of the original supernova \citep{Koo2017}, Cas~A is the only core-collapse remnant with a secure SN classification. Spectroscopy performed on light echoes associated with the original supernova, from multiple lines of sight, found that the supernova was most similar to SNe IIb events SN 1993J and 2003bg near maximum light  \citep{Krause08,Rest08,Rest11}). This makes the remnant very helpful in attempts to connect unresolved SNe with nearby, resolved SNRs \citep{MF17}. 

The bulk of the ejecta of Cas~A are distributed in a ring called the ``main shell'' that emits strongly at X-ray, optical, and infrared wavelengths (Figure~\ref{fig:casa-crab}). These ejecta, having a total mass of $\approx 3$\,M$_{\odot}$, are composed primarily of oxygen, sulfur, argon, neon, and iron \citep{Ennis2006,DeLaney2010,HL12} that has encountered the reverse shock. The ejecta contain considerable amounts of dust ($\approx 0.5$\,M$_{\odot}$; \citealt{DeLooze17}) with varying compositions and sizes \citep{Rho08}. Dust emission is also present throughout the surrounding CSM seen along the periphery of the remnant that has been heated by the forward shock. Approximately $0.5$\,M$_{\odot}$ of ejecta have not yet encountered the reverse shock and emit in the infrared \citep{DeLaney14,LT2020}.

Models for Cas~A suggest that the 15--25 M$_{\odot}$ zero-age-main-sequence progenitor star lost the majority of its mass prior to explosion as a $\approx 4$--6\,M$_{\odot}$ star \citep{CO03,HL12,Lee14}. The mass loss was likely encouraged through interaction with a binary companion \citep{Young06,Sato20}. Deep observations targeting the center of expansion have ruled out the possibility of a surviving companion star for the putative binary progenitor system \citep{Kochanek2018,Kerzendorf2019}, which is in tension with claims of OB companions detected at the locations of extragalactic stripped-envelope explosions \citep{Maund2004,Ryder2018,Fox2022}. If the progenitor system was indeed a binary, the two stars may have merged or were disrupted prior to collapse. More exotic scenarios such as a surviving white dwarf, neutron star, or black hole are unlikely but cannot be ruled out.

Cas A exhibits exceptionally high velocity Si- and S-rich material in a jet/counter-jet arrangement in the NE and SW directions \citep{Fesen01,Hwang04}. The known extent of this jet region contains fragmented knots of debris traveling up to 15000 km/s, which is three times the velocity of the bulk of the O- and S-rich main shell. The large opening half-angle $\approx 40^{\circ}$ is inconsistent with a highly collimated flow \citep{MF13}, and the energy in this region ($\sim 10^{50}$\,erg) is a fraction of the total energy of the original supernova ($\sim 10^{51}$\,erg) that is best understood as being driven by low-mode convective instabilities encouraged by uneven neutrino heating during core collapse \citep{Vink2004,Grefenstette14}. Nonetheless, some jet-like mechanism carved a path allowing interior material from the Si- S-Ar-Ca region near the core out past the mantle and H- and He-rich photosphere \citep{Laming06,FM16}. The NE-SW outflows of Cas~A may be a common feature of core-collapse supernovae, and supports the view that varying levels of participation from a rapidly rotating neutron star may exist in SNe \citep{Mazzali2008,Soderberg2010,Mazzali14,Metzger15,Milisavljevic2018}.

``First-light'' images of Cas A taken by {\sl Chandra} revealed a central X-ray point source \citep{Tananbaum99,pavlov2000}, that is presumably its remnant neutron star.  This neutron star is part of the family of {\it central compact objects} (CCOs), which are young, exhibit X-ray emission that is steady and predominantly thermal, have magnetic fields that are two orders of magnitude lower than those of typical young NS, and lack surrounding pulsar wind nebulae \citep{Pavlov2004}.  Deep imaging with the James Webb Space Telescope did not detect the CCO and place strong constraints on scenarios involving a possible fallback disk \citep{Milisavljevic2024}.

\begin{figure}[tp]
\centering
\includegraphics[width=0.98\linewidth]{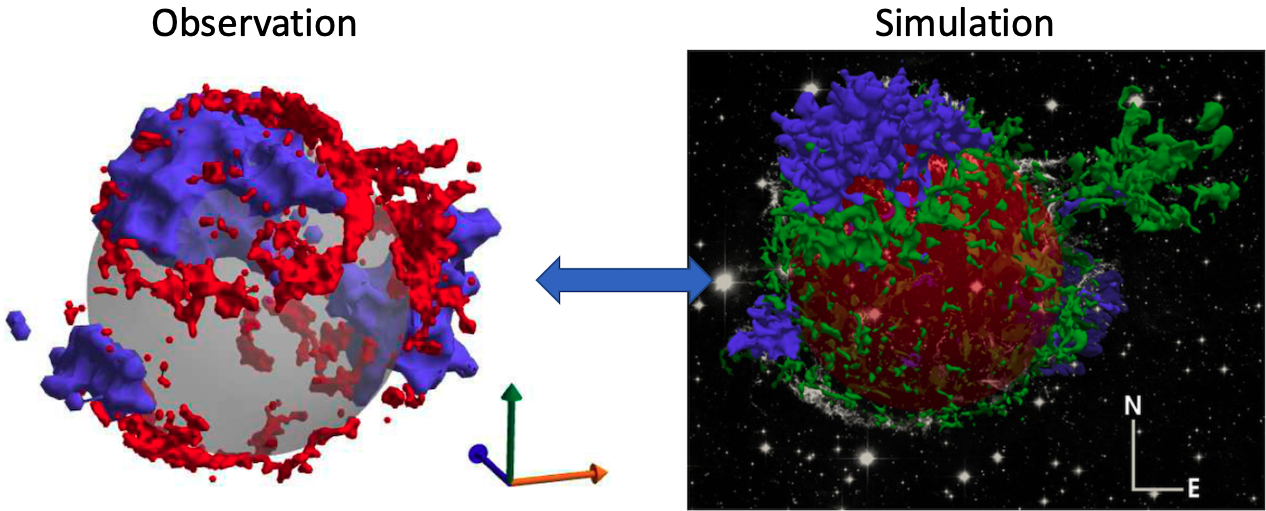}

\caption{Comparison of 3D kinematic reconstruction of Cas A showing optical (red) data tracing S- and O-rich ejecta \citep{Milisavljevic2013} and X-ray (blue) data tracing Fe-rich ejecta \citep{DeLaney2010} to a simulation that evolved a Cas A-like core-collapse explosion to the remnant phase where green traces shocked S- and Si-rich ejecta, and blue again traces shocked Fe-rich ejecta \citep{Orlando2016}. The two viewing perspectives are approximately the same. An interactive web app of observational data is available \href{https://www.physics.purdue.edu/kaboom/casa-webapp/}{here}, and the simulation data can be explored \href{https://sketchfab.com/3d-models/supernova-remnant-cassiopeia-a-99219993de9f4f0e811abd56dce6d288}{here}.}

\label{fig:obs-sim}
\end{figure}

\subsection{Crab Nebula - The prototypical SNR with a pulsar wind nebula}

The Crab Nebula was the first example of a Galactic supernova remnant that was connected with the historical record of a ``guest star'' in 1054 CE \citep{Lundmark1921,Mayall1939}. The event reached an apparent magnitude brighter than $-5$ (associated with absolute magnitude approaching $-18$ mag), was visible during daylight hours for 23 days, and remained visible for approximately two years \citep{Duyvendak1942}. 

Moving from the inside out, the Crab Nebula consists of the Crab pulsar, the synchrotron nebula, and a bright expanding shell of thermal gas in a network of filaments (Figure \ref{fig:casa-crab}).   The Crab pulsar has a period $P = 33$ ms, a period derivative of $\dot{P} = 4.21 \times 10^{-13}$, and spin-down luminosity of $L_{spin}=4\pi^{2}I\frac{\dot{P}}{P^3} \sim 5 \times 10^{38} \rm erg\,s^{-1}$, where $I$ is the moment of inertia of the pulsar ($\approx 1.1 \times 10^{45} \rm g\,cm^{2}$). The majority of the pulsar’s spin-down luminosity is carried away by some combination of magnetic dipole radiation and an ultrarelativistic wind that powers the nebula and is confined by the thermal ejecta into which it expands \citep{Hester2008}. Much of the filamentary ejecta, which are possibly mixed with some material from the pre-supernova wind \citep{Fesen1992}, breaks up into inward pointing fingers of emission due to Rayleigh-Taylor instabilities introduced by PWN pressure. 

The Crab Nebula is an interesting contrast to Cas~A. Although a historical record exists of its light curve, to date no light echoes have been detected and hence no confirmation of its specific core-collapse classification has been possible. It has  expansion velocities ($<$ 2000 km/s ) far lower than commonly seen in SNe, an extremely luminous pulsar, and ejecta showing only helium enrichment.  The total mass of its ejecta is 2-7\,M$_{\odot}$ \citep{Fesen1997,OB2015}, and, as estimated from the small carbon and oxygen abundances in the helium-rich nebula, the plausible mass of the progenitor is 8-13 M$_{\odot}$ \citep{Nomoto87,OB2015}. 

Debate continues as to which SN subtype best matches its properties \citep{DF85,Smith13,YC15}. SN\,1054 was more luminous than a normal Type II SN ($-18$ mag vs.\ $-15.6$ mag), yet its kinetic energy ($\lesssim 10^{50}$\,erg) is surprisingly low compared to the canonical $\sim 10^{51}$\,erg. It was originally speculated that most of the mass and 90\% of the kinetic energy of SN 1054 reside in an invisible freely expanding envelope of cold and neutral ejecta far outside the Crab \citep{Chevalier1977}. However, this fast envelope has never been detected to remarkably low upper limits (\citealt{Lund2012}; but see also \citealt{Hester2008}). \citet{Fesen1997} suggested that SN 1054 was a low energy SN with additional luminosity provided by circumstellar interaction. \citet{Smith13} supports this view and identified potential Crab-like analogs in many recent Type IIn events. \citet{Tominaga13} found that the high peak luminosity could instead be related to the large extent of the progenitor star and not necessarily associated with strong circumstellar interaction.  


Many models originally favored an electron capture SN origin (e.g., \citealt{Nomoto1982} and \citealt{Kitaura06}) in light of the Crab's low energy explosion and unique chemical abundances that included a high Ni/Fe ratio. In this scenario, a massive star whose mass is slightly smaller than the mass required to form an Fe core can still make an electron degenerate O+Ne+Mg core, which can trigger core collapse through electron-capture reactions. However,  this explanation comes with many discrepancies. Electron capture explosions are generally faint ($M_V$ $> -15$ mag) and thus inconsistent with the relatively luminous SN unless SN-CSM interaction was involved, and the proper motion of the pulsar in the Crab Nebula has an associated velocity ($\approx 160$\,km\,s$^{-1}$; \citealt{Kaplan2008}) that is too high to originate from an electron capture explosion \citep{Stockinger2020}. Modern simulations can now produce core-collapse SN explosion of similar low energies, and the latest measurement of the Ni/Fe abundance ratios in the Crab are consistent with those expected from a low-mass Fe core-collapse SN \citep{Temim2024}.





\subsection{3D morphology and modeling SN-SNR evolution}

Nearby, young SNRs can be kinematically mapped via scanning of long slit spectroscopy, widefield integral field unit spectrographcs (e.g., MUSE on the Very Large Telescope; SITELLE on Canada-France-Hawaii Telescope), or Fabry-Perot imaging spectroscopy. If the ejecta motion and center of expansion are known via  2D proper motion studies, and trajectories are assumed to be ballistic, a three dimensional reconstruction can be created by converting the radial velocities of the expanding gas to a z-space coordinate from the plane. To date, kinematic maps of SNRs have been made for Cas A \citep{DeLaney2010,MF15}, the Crab Nebula \citep{Martin2021}, N132D \citep{Law2020}, 0540-69.3 \citep{Larsson2021}, 1E\,0102.2-7219 \citep{Vogt2018}, and SN\,1987A \citep{Larsson2016}.

\begin{figure}[tp]
\centering
\includegraphics[width=0.95\linewidth]{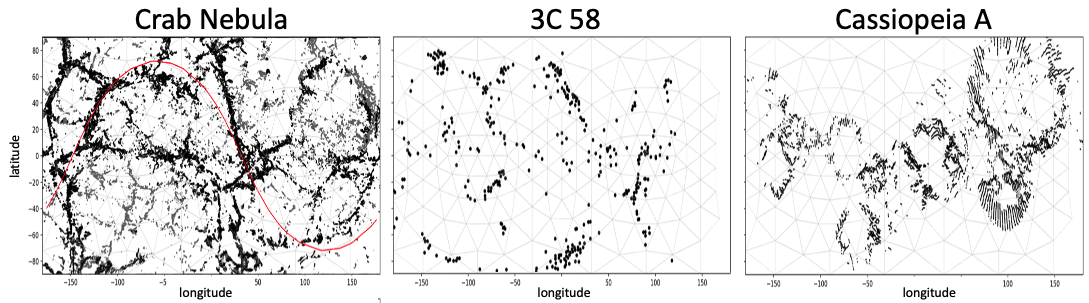}

\caption{Mercator projections of optically-emitting material represented in the 3D reconstructions of the Crab Nebula \citep{Martin2021}, 3C\,58 \citep{LF18} and Cas A \citep{MF13}. For the Crab, the densest and deepest material is represented in black, while the faster expanding material is in grey. The red line shows the pulsar torus plane as fitted by \citet{NR04}. Adapted from \citet{Martin2021}.}
\vspace*{-3mm}
\label{fig:mercator}
\end{figure}

Figure~\ref{fig:obs-sim} shows a detailed 3D reconstruction of Cas~A, and Figure~\ref{fig:mercator} shows a comparison of the Mercator projections of Cas A \citep{MF13}, the Crab \citep{Martin2021}, 3C58 \citep{LF18}.  The 3D kinematic reconstruction of Cas A shows that its ejecta are arranged in several well-defined and nearly circular rings with diameters between approximately 30$^{\prime\prime}$ (0.5 pc) and 2$^{\prime}$ (2 pc), demonstrating that the distribution of its metal-rich ejecta is not random. The sizes and arrangement of the large-scale rings reflect properties of the explosion dynamics and subsequent evolution of the expanding debris. Three large concentrations of Fe-rich ejecta, which are good tracers of the original explosion dynamics, lie within and bounded by ring structures. This complementary arrangement is consistent with the ``Ni bubble effect'' having influenced the remnant's expansion dynamics shortly after the original explosion, wherein the observed ejecta rings represent cross-sections of large cavities in the expanding ejecta created by a post-explosion input of energy from plumes of radioactive Ni-rich ejecta. Nickel bubbles may be a common phenomenon of young, core-collapse SNRs \citep{Li1993,Milisavljevic2012}

The Crab exhibits a much more complex structure compared to Cas A. The largest and deepest structures of the Crab are similar both in size and shape as those seen in 3C 58. In these cases, Rayleigh-Taylor instabilities encouraged by the PWN are likely dominating the morphology of the ejecta. The general boundaries of the 3D volume occupied by the Crab are not strictly ellipsoidal as commonly assumed, and instead appear to follow a ``heart-shaped'' distribution that is symmetrical about the plane of the pulsar wind torus. Conspicuous restrictions in the bulk distribution of gas consistent with constrained expansion coincide with positions of the dark bays and east–west band of He-rich filaments, which may be associated with interaction with a pre-existing circumstellar disc. The distribution of filaments follows an intricate honeycomb-like arrangement with straight and rounded boundaries at large and small scales that are anticorrelated with distance from the centre of expansion. Comparing the observed morphology to the recent simulations of \citet{Stockinger2020}, \citet{Martin2021}  disfavoured associating SN\,1054 with an electron capture supernova and instead with an Fe-core progenitor with less second dredge-up.

Pristine structures and features recovered in resolved SNRs reflect fundamental properties of the explosion dynamics and the early interaction of the SN blast wave with the inhomogeneous CSM. Hence, these kinematic maps provide unique opportunities to test the origins of turbulent mixing and explosion asymmetry predicted in simulations. The ability to link the physical and chemical characteristics of core-collapse SNRs to the explosion processes associated with their parent SNe through hydrodynamic / magnetohydrodynamic (MHD) models has been improving over the last decade (Figure~\ref{fig:obs-sim}).  Modeling the SN--SNR connection involves many factors, including: the structure and chemical stratification of the stellar progenitor at collapse, the explosive nucleosynthetic processes, the effects of post-explosion anisotropies, the dynamics and chemical evolution of ejecta since the SN, and the interaction of the SNR with the surrounding ambient medium \citep{Orlando2020,Orlando2021}.

\section{Summary and outlook}
\label{sec:summary}

\bluee{The theory of the supernova explosion mechanism (Sec. \ref{sec:bernhard}), the study of supernova light curves and spectra (Sec. \ref{sec:lc_spec}), and the study of supernova remnants (Sec. \ref{sec:remnants}) have always fertilised each other, and recent progress in these fields has further improved the prospects for this. As with any scientific field, progress arises over iterations in a loop between models and data, with both forward (processing hydrodynamic simulations with radiative transfer codes) and backward modelling (analysing data with analytic methods) playing important roles. The full ab-initio forward modelling is challenged by the enormous computing costs involved in simulating the explosion and evolution of supernovae - the ejecta needs to be evolved to times of weeks or months, but even the first second of core-collapse can take millions of CPU hours. There is therefore a vast range of methods in the range between ``full ab-initio'' and ``toy models'' - the challenge is to patch and sort the results from these into a coherent and consistent framework. As spatially resolved SN remnants clearly show us - SNe are extraordinarily complex objects, and this task is therefore a grand challenge.}

\bluee{We consider it fair to say that simulations are now in a position to inform us at quite some depth about the physical principles behind CCSN explosions, and also to allow for a quantitative understanding of some of the observed properties of SNe and their compact remnants. Admittedly, this relies to a large extent still on \emph{phenomenological} models
that require calibration,
but the capabilities of fully self-consistent models are growing rapidly.}




\bluee{A key question that simulations have started to address is how the progenitor mass, metallicity,  rotation, and possibly binarity,  determine whether a massive star explodes as a CCSN or collapses into a BH.
There is significant evidence from identified supernova progenitors in archival images that there is
a dearth of explosions from red supergiants above progenitor
masses of $M_{ZAMS}\sim 18\   M_\odot$ \citep{Smartt2015}. Three-dimensional modelling of supernova explosions is not yet at the stage where the mass range of successful explosions can be confidently predicted. However, parameterised models of neutrino-driven explosions \citep{mueller_16,sukhbold_16} can account for the absence of explosions at high masses and suggest that, despite model uncertainties, inherent characteristics of the progenitor star strongly influence explodability, such as mass size of the iron-silicon core, whose edge defines a natural point for the initiation of an explosion. An important research avenue for this question is also the ongoing monitoring of millions of red supergiants, waiting to see if some of them disappear without exploding as a SN \citep{Kochanek2008}.}

\bluee{Furthermore, the physics revealed by simulations of neutrino-driven explosions can explain some of the systematics
of observed explosion and remnant properties \citep[e.g.][]{Fryer2012}. 
Since longer accretion onto the proto-neutron star implies
more energetic explosions, one naturally expects a correlation between the progenitor structure and the explosion energy in the neutrino-driven paradigm.
Progenitors with more massive C/O cores tend to accrete for longer and hence reach higher explosion energies.
Population studies using calibrated phenomenological explosion models suggest that neutrino-driven explosions can thus produce a range of explosion energies from $\sim 10^{50}\,\mathrm{erg}$ to $\sim 2\times 10^{51}\,\mathrm{erg}$, which are loosely correlated with progenitor mass \citep{Mueller2016}. A correlation between the $^{56}$Ni mass and the ejecta mass is also theoretically expected, quite regardless of the explosion mechanism, because higher explosion energies permit substantially more explosive burning to the iron-group in the innermost shells of the ejecta due to the shock managing to heat a larger region over the Si-burn limit ($5 \times 10^9$ K). In the neutrino-driven mechanism, the neutrino-heated ejecta may also produce substantial $^{56}\mathrm{Ni}$ by nucleon recombination. Models of neutrino-driven explosions are in general good agreement with the observed range of nickel masses of up to $\mathord{\sim} 0.15\  M_\odot$ - the $^{56}$Ni mass is arguably the most robustly inferred quantity from observations.}

\bluee{In contrast to this, it has turned out quite difficult to determine unique values for $E$ and $M_{\rm ej}$; for Type IIP SNe there are several parameter degeneracies at play, e.g., velocity information not breaking degeneracy with $R_0$ \citep{Goldberg2019}, and the plateau light curve not being sensitive to the total mass but just the envelope mass  \citep{Dessart2019}. Parameter degeneracies may be mitigated by including the early shock cooling phase, but this in turn has turned out to be sensitive to yet further parameters related to the CSM \citep{Morozova2018}, as well as to the detailed 3D structure of the outermost material. For SESNe it is challenging to accurately infer $E$ because the diffusion time has only a weak dependency on it ($\propto E^{-1/4}$) and the peak luminosity none (depending rather on the $^{56}$Ni mass). 
The timing of gamma-ray escape in the tail phase, having a stronger $E$ dependency than the diffusion time ($t_{trap} \propto E^{-1/2}$), can be of use to break the degeneracy. A completely different way to estimate  $E$ can be done the cases when the SN runs into a massive CSM (luminous Type IIn SNe), turning $\gtrsim$ 50\% of the kinetic energy into radiation - a handful of cases here have given $E \gtrsim 0.5\ \rm{B}$ \citep[e.g][]{Fransson2014}. Even here there can be caveats, as it is sometimes difficult to distinguish a CCSN and thermonuclear SN origin for Type IIn SNe \citep{Jerkstrand2020}.}



\subsection{Outlook}
As we look to the future, there is reason for optimism for exciting progress in the years ahead. Automated transient surveys have revolutionzed the field of supernovae over the past 20 years, and with the Legacy Survey of Space and Time soon starting up at the Vera C. Rubin Observatory there will be yet another leap forward in the number of detected transients. While in the past there were so few SNe detected that one followed them all, we now have to be very selective. This means an early entry of theory and simulations into the process, for providing inputs into what is of most relevance to follow up on, and with which  instrumentation. The improvements in detection efficiencies will also, with the upcoming generation of Extremely Large Telescopes (ELTs), be matched with improvements in the long-term monitoring data we can obtain. The emphasis both with the ELTs and the James Webb Space Telescope is on the infrared - and indeed this defines much of the vista to be explored in the coming years, with rich expected signatures of    nucleosynthesis, molecules, and dust.

It is our view that future progress in SN research lies in combining models and data covering multiple phases and multiple wavelength windows. 
In this new multi-messenger era, observations of neutrinos \citep{mueller_arnps} and gravitational waves \citep{abdikamalov_22,Vartanyan2023} are also anticipated to contribute tremendously to a better understanding of CCSNe in the event of an explosion in the Milky Way or its close neighborhood. As many sections in this review has emphasized, any limited patch of data and models typically does not give unique solutions and answers -  the grand challenge lies with combining the patches and doing a holistic analysis. In that sense, SN research by necessity will continue to transition from individual or small-group efforts into larger, global collaborative ones - and this is all and well, in the true spirit of science.

\begin{ack}[Acknowledgments]

The authors thank S. Valenti for support in sourcing the data and putting together Figure 7, and K.\ Maguire for sourcing the data for Figure 2. We thank M.\ Modjaz, K.\ Maeda, S.\ Valenti, T.\ Foglizzo, J.\ Goldberg, and T.\ Temim for discussion and comments on the manuscript.  
A.J.\ acknowledges support by the European Research Council, the Swedish National Research Council, and the Knut and Alice Wallenberg foundation. D.M.\ acknowledges NSF support from grants PHY-2209451 and AST-2206532. 
B.M.\ acknowledge support by
the Australian Research Council (ARC)
through grants FT160100035 and DP240101786 (BM, AH),
by Australia Limited's ASTAC scheme, the National Computational Merit Allocation Scheme (NCMAS), and by an
Australasian Leadership Computing Grant (ALCG).

\end{ack}

\seealso{Nebular spectra; \citet{Jerkstrand2017}. Classification; \citet{GalYam2017}. SN neutrino emission; \cite{Janka2017}. 
Dust formation in SNe; \cite{Sarangi2018}. Supernova remnants; \citet{Vink2020}. Supernova-to-supernova remnant transition; \citet{MF17}.}

\bibliographystyle{Harvard}
\bibliography{reference}

\end{document}